# Moving bits instead of watts: Accelerating global AI data-centre deployment at the energy-digital-water-land nexus

**Jiawei Qu[1,2,3], Xiaoying Shi[1,3], Kai Hou[2]*, Hongjie Jia[2], Jinyue Yan[1,3]***

[1] *Department of Building Environment and Energy Engineering, The Hong Kong Polytechnic University; Hong Kong, SAR, China.*

[2] *State Key Laboratory of Smart Power Distribution Equipment and System, Tianjin University; Tianjin, China.*

[3] *International Centre of Urban Energy Nexus, The Hong Kong Polytechnic University; Hong Kong, SAR, China.*

E-mail: * khou@tju.edu.cn, j-jerry.yan@polyu.edu.hk

**Acknowledgments**

1) The Hong Kong Polytechnic University project P0047700, International Centre of Urban Energy Nexus (JQ, XS, JY)
2) The Hong Kong Polytechnic University project P0056532, IPT4U: Intelligent Platform & Toolbox for Urban Infrastructure Resilience (JQ, XS, JY)
3) National Science and Technology Major Project, Low-Carbon and High-Reliability Urban Distribution Power Systems, grant 2024ZD0800700 (JQ, KH, HJ)

**Highlights:**

1) Global data-centre siting optimization to alleviate grid-connection bottlenecks
2) Data-centre construction cost quantification over 4,000 subnational regions
3) Energy-water-fibre-land coupling in global data-centre siting optimization
4) Global energy-system modelling with over 240,000 transmission lines

# Moving bits instead of watts: Accelerating global AI data-centre deployment at the energy-digital-water-land nexus

**Jiawei Qu[1,2,3], Xiaoying Shi[1,3], Kai Hou[2]*, Hongjie Jia[2], Jinyue Yan[1,3]***

[1] *Department of Building Environment and Energy Engineering, The Hong Kong Polytechnic University; Hong Kong, SAR, China.*

[2] *State Key Laboratory of Smart Power Distribution Equipment and System, Tianjin University; Tianjin, China.*

[3] *International Centre of Urban Energy Nexus, The Hong Kong Polytechnic University; Hong Kong, SAR, China.*

**Abstract:** Artificial intelligence is driving an unprecedented expansion of data centres. However, this growth is increasingly stalled by grid transmission constraints that trap new data centres in years-long connection queues. Unlike electricity, computation can be moved across the globe through fibre-optic networks, but whether this spatial transfer can bypass constrained grids and accelerate global AI data-centre deployment remains unknown. Here we develop a global framework that places future data-centre capacity in energy-abundant regions and routes computational workloads through fibre-optic networks rather than transmitting electricity. We evaluate the framework through more than 11 million hourly operation simulations at subnational resolution, integrating constraints on energy systems, water, digital connectivity, land and financing risk. Results show that across multiple data-centre development and climate-policy scenarios, leveraging global data transfer allows new data centres to be sited where these conditions are favourable, lowering construction costs by 44-47% and carbon emissions by 38%. By treating the global fibre-optic network as an energy-system asset, it avoids 34-67 GW of high-voltage transmission capacity, comparable in magnitude to Italy's national peak electricity demand. Crucially, this methodology could bring more than 95% of connection-delayed data centres online in 2030, with Canada, the Nordic countries and the Arabian Peninsula emerging as preferred locations for data-centre development. Our findings establish computational mobility as an alternative to energy-system expansion, informing the coordinated global development of energy and artificial-intelligence infrastructure.

***Nomenclature***

| | | | |
|---|---|---|---|
| *Indices and Sets* | | | |
| $r$ | index of the locations | $l$ | index of transmission lines |
| $N$ | lifetime of data centres | $p$ | index of land-cover pixels |
| $m$ | index of financing models | $g$ | index of electricity-generation technologies |
| $t$ | index of time | $G_r$ | set of fibre-connected locations accessible from location $r$, without latency restrictions |
| $i,j$ | index of origin and destination locations for computational workload transfer | $N_r$ | set of locations accessible from location $r$ that satisfy latency constraints |
| *Parameters and Constants* | | | |
| $u_r^{\mathrm{GS}}$ | unit construction cost of grid-supplied data centres at location $r$ | $u_r^{\mathrm{SG}}$ | unit construction cost of self-generation data centres at location $r$ |
| $u_r^{\mathrm{gc}}$ | unit cost of general requirements and general contractor fees for data-centre construction at location $r$ | $c_g^{\mathrm{gen}}$ | additional unit construction cost attributable to onsite generation technology g at location $r$ |
| | | $c_r^{\mathrm{land}}$ | industrial land price at location $r$ |
| $u_r^{\mathrm{shell}}$ | Unit construction cost of the data-centre building shell at location r | $c_{g,r}^{\mathrm{gen}}$ | unit capital investment cost of onsite generation technology g at location $r$ |
| $u_r^{\mathrm{mech}}$ | unit cost of mechanical and cooling systems for data centres at location $r$ | $A_{g,r}^{\mathrm{land}}$ | land area required per unit of installed capacity of onsite generation technology g at location $r$ |
| $u_r^{\mathrm{elec}}$ | unit cost of electrical and backup-power systems for data centres at location $r$ | $\lambda^{\mathrm{DC}}$ | penalty cost per unit of data-centre IT capacity remaining unconnected |
| $\rho_{r,m}$ | annual discount rate at location $r$ under financing mode m. | $\phi_{r,t}^{\mathrm{wnd}}$ | hourly wind capacity factors |
| | | $D_{r,t}^{\mathrm{DC}}$ | data-centre IT power demand at location $r$ |
| $\omega_g$ | water-use intensity of onsite power-generation technology $g$ | $c_{r,g}^{\mathrm{inv}}$ | unit investment cost of onsite power-generation technology $g$ at location r. |
| $\alpha^{\mathrm{DC}}$ | fraction of fully flexible data-centre workloads eligible for temporal or latency-unconstrained spatial shifting | $\Delta K_l^{\mathrm{tr}}$ | planned transmission capacity of transmission line $l$ |
| | | $c_{r,g}^{\mathrm{FOM}}$ | unit fixed operation and maintenance cost of onsite generation $g$ |
| $\phi_r^{\mathrm{DC}}$ | demand curve for data-centre workloads | $a_p$ | area of land-cover pixel $p$ |
| $\eta^{\mathrm{st}}$ | charging or discharging factor of the energy-storage | $F_l^{\mathrm{ex}}$ | existing power-transfer capacity of transmission line $l$ |
| $\alpha^{\mathrm{load}}$ | temporal flexibility factor for conventional load | $\phi_{r,t}^{\mathrm{sol}}$ | hourly solar capacity factors |
| $\eta^{\mathrm{ch}}$ | charging efficiency of the energy-storage | $B_{r,l}$ | bus-branch incidence matrix |
| $\eta^{\mathrm{dis}}$ | discharging efficiency of the energy-storage | $\mathrm{PUE}_r$ | the power usage effectiveness of data centres |
| $\underline{s}$ | minimum state-of-charge of the energy-storage | $\mathrm{WUE}_r$ | the water usage effectiveness of data centres |
| $\overline{s}$ | maximum state-of-charge of the energy-storage | $W_r^{\mathrm{avail}}$ | freshwater availability budget at location $r$ |
| $\alpha_r$ | the fraction of water available for industrial use | $c_r^{\mathrm{grid}}$ | unit cost of grid electricity purchased |
| $c_r^{\mathrm{gas,conn}}$ | unit investment cost of constructing a new natural-gas pipeline connection | $d_r^{\mathrm{gas}}$ | distance from location $r$ to the nearest accessible natural gas network |
| $c_r^{\mathrm{fi}}$ | investment cost of connecting location $r$ to the nearest accessible fibre-optic network. | $c_r^{\mathrm{gas,stn}}$ | investment cost of constructing a natural-gas metering station at location $r$ |
| $c_r^{\mathrm{node}}$ | investment cost of establishing a new Internet exchange point at location $r$ | $c^{\mathrm{fi}}$ | unit investment cost of constructing new fibre-optic network connections |
| $lat$ | round-trip propagation delay per unit distance along a fibre-optic network. | $z_r^{\mathrm{fi}}$ | distance from location r to the nearest accessible fibre-optic network |
| $L_s^{\mathrm{proc}}$ | processing delay | $\alpha$ | Fibre-route inflation factor |
| $M_p$ | binary indicator of land suitability for power-generation development in pixel $p$ | $\overline{K}_{r,\mathrm{wnd}}^{\mathrm{self,gen}}$ | maximum installable onsite wind-power generation capacity at location $r$ |
| $\overline{K}_{r,\mathrm{sol}}^{\mathrm{self,gen}}$ | maximum installable onsite solar-power generation capacity at location $r$ | $c_{r,g}^{\mathrm{VOM}}$ | unit variable operation and maintenance cost of onsite generation $g$ |
| $K_r^{\mathrm{DC}}$ | data-centre IT capacity at location $r$ in 2025 | $K_r^{\mathrm{DC,demand}}$ | data-centre projected IT capacity at location $r$ in 2030 |
| *Variables* | | | |
| $K_{r,g}^{\mathrm{self,gen}}$ | installed capacity of onsite generation | $K_r^{\mathrm{DCgrid}}$ | installed IT capacity of grid-supplied data centres |
| $C^{\mathrm{tot}}$ | total annualized costs | $K_r^{\mathrm{DCself}}$ | installed IT capacity of self-generation data centres |
| $K_r^{\mathrm{DC,uns}}$ | data-centre IT capacity remaining unconnected | $C_r^{\mathrm{land}}$ | total land cost associated with onsite generation |
| $G_{r,t}^{grid}$ | power supplied by the public grid to data centres | $C_r^{\mathrm{DC}}$ | data-centre construction costs |
| $C_r^{\mathrm{opr}}$ | operating cost for data centres | $L_{r,t}^{\mathrm{up}}$ | upward adjustment variables for conventional load |
| $G_{r,g,t}^{\mathrm{self,gen}}$ | power output of onsite generation | $G_{r,\mathrm{wnd},t}^{\mathrm{self,gen}}$ | power output from onsite wind generation |
| $L_{r,t}^{\mathrm{dn}}$ | downward adjustment variables for conventional load | $D_{r,t}^{\mathrm{sp,up}}$ | upward adjustment to data-centre IT load through latency-constrained spatial workload transfer |
| $D_{r,t}^{\mathrm{te,up}}$ | Upward adjustment to data-centre IT load through temporal workload shifting | $D_{r,t}^{\mathrm{sp,dn}}$ | downward adjustment to data-centre IT load through latency-constrained spatial workload transfer |
| $D_{r,t}^{\mathrm{te,dn}}$ | downward adjustment to data-centre IT load through temporal workload shifting | $C^{\mathrm{gas,net}}$ | total cost of natural-gas network interconnections and associated metering stations |
| $D_{r,t}^{\mathrm{net}}$ | net data-centre IT load | $G_{r,\mathrm{sol},t}^{\mathrm{self,gen}}$ | power output from onsite solar generation |
| $L_{r,t}^{\mathrm{net}}$ | net conventional load | $K_r^{\mathrm{st}}$ | installed capacity of the energy-storage system |
| $P_{r,t}^{\mathrm{ch,self/grid}}$ | energy-storage charging power | $S_{r,t}$ | energy-storage state of charge |
| $P_{r,t}^{\mathrm{dis,self/grid}}$ | energy-storage discharging power | $F_{l,t}$ | active-power flow on transmission line $l$ |
| $D_{r,t}^{\mathrm{sp,as,up}}$ | upward adjustment to data-centre IT load through latency-unconstrained spatial workload transfer | $D_{r,t}^{\mathrm{sp,as,dn}}$ | downward adjustment to data-centre IT load through latency-unconstrained spatial workload transfer |
| $F_{ir,t}^{as}$ | IT workload transferred from location $i$ to location $r$ without latency constraints | $F_{ir,t}$ | IT workload transferred from location $i$ to location $r$ subject to latency constraints |
| $D_{r,t}^{\mathrm{net,grid}}$ | net IT load of grid-supplied data centres | $D_{r,t}^{\mathrm{net,self}}$ | net IT load of self-generation data centres |

## 1. Introduction

Artificial intelligence (AI) has rapidly become a general-purpose technology reshaping science, industry and society [1],[2]. This progress has driven a rapid growth in computational demand that is highly concentrated geographically [3]. In 2024, data centres consumed approximately 415 TWh of electricity, equivalent to about 1.5% of global demand [4], and this is projected to reach 945 TWh by 2030 [4]. This demand is highly concentrated: the United States, China, and Europe account for approximately 45%, 25% and 15% of global data-centre electricity use, respectively [4], with Northern Virginia alone hosting roughly 13% of operational capacity worldwide [5], [6]. This spatial concentration has turned local energy supply into a critical bottleneck for data-centre expansion [7], [8]. In energy-constrained regions, hyperscale data centres can typically be commissioned within 1-2 years [4],[9], whereas new transmission lines require 4-8 years to permit and build [4], [10] , and gas-turbine generation now faces multi-year procurement delays for critical components [11]-[13]. As a result, more than 20% of planned data-centre capacity globally is at risk of connection delay [4], with current wait times of 5-7 years in Northern Virginia [6], [14] and reported queue waits exceeding a decade in the United Kingdom [15]. Therefore, energy-system expansion is increasingly struggling to keep pace with AI-driven data-centre growth [16], [17].

One promising pathway is to site data-center capacity where energy is abundant and transfer the computational workloads through fibre-optic networks, rather than transmitting electricity to existing computing hubs. Electricity demand is tied to where computation is performed, but the computational workloads that create that demand can be routed globally through fibre networks while satisfying latency requirements. Data transmission consumes about 0.06 kWh per GB [18], [19], far less than the line losses incurred in delivering the equivalent electricity over high-voltage lines [20]. At the same time, currently lit subsea-cable capacity accounts for only about 50% of what has been deployed [21], suggesting substantial latent capacity in the global fibre networks. In principle, this opens the possibility of decoupling where data-centre capacity is built from where its computational demand originates: siting new data centres in regions with favourable energy, digital connectivity, water and land conditions, and routing workloads accordingly [22]-[25]. Previous studies have examined data-centre energy demand [3], [26]-[28], water use [5], digital connectivity [29] and land [30], alongside spatial workload shifting for carbon optimization [31], [33] and energy management [32], [34]. However, how these interdependent resource constraints jointly shape the feasibility and grid-connection benefits of computation transfer remains poorly understood at the global scale.

To address these gaps, we develop a global computation-transfer-based siting framework that places new data centres in regions where energy, digital connectivity, water and land are favourable, with computational demand allocated accordingly. We first estimate data-centre construction costs across more than 4,000 subnational regions worldwide by integrating these four factors with climate-adapted cooling technologies optimized for power-usage effectiveness (PUE) and water-usage effectiveness (WUE). These subnational costs are then embedded in a siting model coupling global power systems, comprising more than 240,000 high-voltage transmission lines, with global natural-gas and fibre-optic networks across 149 countries. To establish whether the decoupling holds across plausible futures rather than under a single projection, we evaluate the framework under four IEA-derived data-centre development scenarios [4] and nine International Institute for Applied Systems Analysis (IIASA)-derived climate-policy scenarios [35], complemented by a suite of uncertainty analyses. The main contributions of this paper are as follows,

1) A high-resolution framework is developed to quantify construction costs for both grid-supplied and self-generation data centres. The framework captures spatial cost heterogeneity across more than 4,000 subnational regions worldwide, providing an economic basis for optimal data-centre siting.

2) A global computation-transfer-based siting model is proposed to jointly optimize data-centre locations and computational workload allocation. By shifting computing demand towards regions with favourable energy and digital infrastructure, the model provides an alternative to conventional grid-expansion-dependent planning.

Drawing on more than 11 million dispatch simulations across subnational regions worldwide, our analysis yields three critical findings:

1) Unlocking delayed capacity. Under continued local energy-system expansion, 10-15% of data centres worldwide are projected to remain connection-delayed by 2030, whereas computation-transfer-based siting could bring more than 95% of this delayed capacity online.

2) Reducing cost and carbon. By leveraging global data transfer, new data centres can be built in regions with favourable energy, water, fibre and land conditions, lowering construction costs by 44-47%, energy cost by 17.3-20.4% and carbon emissions by 38% compared with energy-system expansion.

3) Avoiding energy expansion. By treating the global fibre-optic network as an energy-system asset, 34-67 GW of additional transmission capacity is avoided, comparable in magnitude to Italy's national peak electricity demand.

The remainder of this paper is organized as follows. Section 2 presents the methodology, including data inputs, quantification of global data-centre construction costs, and the proposed computation-transfer-based siting model. Section 3 presents the global analysis, and Section 4 concludes the paper.

## 2. Methodology

We develop a global computation-transfer-based siting framework comprising three components: data inputs, construction-cost quantification and computation-transfer-based siting model (Fig. 1). In the data-input module, we first compile the locations and capacities of 12,445 data-centre clusters worldwide to establish the 2025 global baseline. Future deployment is then projected under the four data-centre scenarios from the International Energy Agency (IEA). For each region, we integrate region-specific information across four layers: energy, water, digital connectivity and land. The energy layer incorporates global natural-gas networks and power systems, power-system reliability and energy-cost parameters. The water layer combines industrial freshwater withdrawals with technology-specific water-use intensities to constrain water demand for data-centre cooling and onsite generation. The digital connectivity layer includes terrestrial fibre networks, submarine cable systems and internet exchange points, which jointly determine regional network access and latency constraints. The land layer accounts for land availability, renewable-energy potential and industrial land prices.

We then quantify construction costs for data-centre deployment across more than 4,000 subnational regions worldwide. The module combines four resource layers with two additional inputs: (i) deployment mode, including grid-supplied and self-generation options; and (ii) financing mode, including local and multinational financing. To account for different ownership structures, local financing applies country-specific weighted average costs of capital, whereas multinational financing combines a global AI-industry baseline cost of capital with country-specific risk premiums. Regions facing high capital-recovery risk are excluded from the candidate locations.

Finally, we develop a computation-transfer-based siting model to reallocate future data-centre capacity from energy-constrained regions to feasible receiving regions through computation transfer, while satisfying latency, energy, water and land constraints. We evaluate the framework's feasibility, cost-effectiveness and environmental impact across four IEA data-centre development scenarios and nine IIASA climate-policy scenarios.

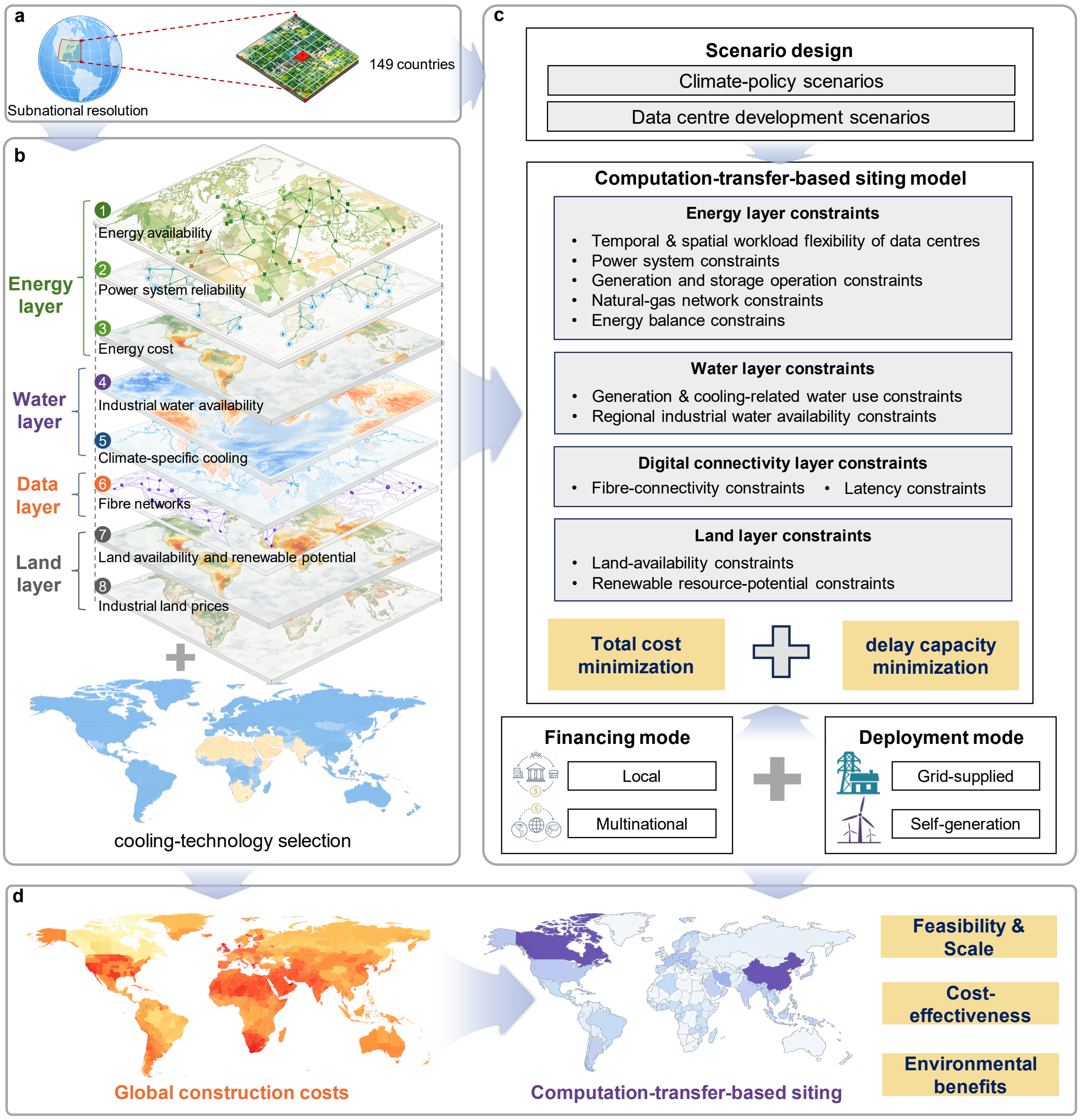

**Fig. 1 Computation-transfer-based siting framework. a,** Spatial resolution of the framework. Construction costs are estimated globally at subnational resolution, while computation-transfer-based siting model covers 149 countries, with subnational detail for major data-centre markets. **b**, Data inputs and construction-cost quantification. The framework integrates energy, digital-connectivity, water and land constraints with cooling-technology selection to estimate spatially explicit construction costs. **c**, Computation-transfer-based siting model. Climate-policy and data-centre development scenarios drive an optimization model that minimizes total costs and delayed capacity across grid-supplied and self-generation deployment modes, and local and multinational financing modes. **d**, Model outputs.

## *2.1. Data inputs*

### *2.1.1 Spatial and temporal scope*

We divide the world into nine regions: Europe (EUR), North America (NA), the Middle East and North Africa (MENA), Oceania (OCE), sub-Saharan Africa (SSA), East and Southeast Asia (ESEA), West Asia (WA), South and Central Asia (SCA) and Latin America and the Caribbean (LAC). Within these regions, we further define more than 4,000 subnational regions using the Natural Earth Admin-1 States and Provinces dataset [36], which represents first-order administrative units below the national level. Data-centre construction costs are estimated at this subnational scale. We further focus on several major data-centre countries, selected according to existing data-centre capacity and projected demand growth: Australia (AU), Brazil (BR), China (CN), Denmark (DK), Finland (FI), France (FR), Germany (DE), Greece (GR), India (IN), Ireland (IE), Italy (IT), Japan (JP), Luxembourg (LU), Mexico (MX), Norway (NO), Poland (PL), Portugal (PT), Russia (RU), South Africa (ZA), South Korea (KR), Spain (ES), Sweden (SE), the United Kingdom (UK) and the United States (US). Temporally, construction-cost estimation and computation-transfer-based siting are conducted for 2030.

#### *2.1.2 Data centres*

Locations and capacities of existing data centres are obtained from DataCentreMap [37], which reports 12,445 data-centre clusters worldwide. Future data-centre capacity follows four IEA scenarios: Headwinds, High Efficiency, Base Case and Lift-Off [4]. These scenarios represent different combinations of artificial-intelligence adoption, efficiency improvement and deployment constraints. To represent operational heterogeneity, data centres are further classified into ten representative space-conditioning technologies spanning large-scale, midsize and small facilities, together with different cooling-system configurations. For each technology, spatially PUE and WUE are estimated from hourly dry-bulb temperature, relative humidity and atmospheric pressure to capture regional variation in cooling performance and water demand (see Supplementary Text 1).

#### *2.1.3 Energy layer*

The energy layer includes energy availability, power-system reliability and energy costs. Energy availability constrains access to global power systems and natural-gas networks; power-system reliability determines backup and redundancy requirements; and energy costs determine the cost of grid-supplied electricity.

1) Energy availability

We represent energy infrastructure using global power-system and natural-gas network data. We focus on the major data-centre countries, where deployment scale, connection pressure and grid constraints are most relevant to the global siting problem. For major countries, we reconstruct their transmission networks from OpenStreetMap [38], retaining lines at 220 kV and above. For all other countries, we aggregate the power system into a single national node. The resulting network contains more than 240,000 high-voltage transmission lines globally. Importantly, this topology is not intended to build a global power-flow model, but to identify where power constraints may delay the connection of new data centres. Future national generation shares for major technologies are derived from the IIASA scenario database [35]. For major countries, we compile subnational demand and generation mixes from electricity yearbooks, annual power-sector reports and official statistics, resolving the shares of bioenergy, coal, gas, hydropower, nuclear, solar, wind and oil generation across subnational regions. Global hourly electricity demand profiles are generated using Renewables.ninja [69]. Transmission expansion plans in the major countries are compiled from government plans, power-system datasets and previous transmission expansion studies. Given that onsite gas-fired generation depends on physical access to fuel supply, candidate locations are further constrained by their proximity to existing gas infrastructure or the need to build new gas-network connections. We therefore incorporate the global natural-gas transmission pipelines and LNG terminals, characterizing region natural-gas accessibility [39].

2) Power system reliability

Power system reliability is represented by using the system average interruption duration index (SAIDI) and the system average interruption frequency index (SAIFI). SAIDI measures the average annual outage duration experienced by customers, whereas SAIFI measures the average annual number of interruptions. National values are obtained from global reliability datasets [40]. For the major countries, subnational reliability statistics are used where available, including state-level data for the United States [41], Australia [42], Europe [43] and major Asia-Pacific economies [44]. Where subnational data are unavailable, national reliability index are downscaled using a nonlinear GDP-based scaling method to capture within-country variation in grid reliability.

3) Energy costs

We compile investment costs, fixed operation and maintenance costs, and variable operation and maintenance costs for bioenergy, coal, gas, hydropower, nuclear, solar, wind and oil-based generation technologies. These parameters are drawn from the IEA, the U.S. Energy Information Administration, government reports and previous studies.

### *2.1.4 Water layer*

To represent existing industrial freshwater demand, we use national industrial freshwater-withdrawal data from FAO AQUASTAT [45], [46] and checked against the Our World in Data compilation [47]. Industrial withdrawals are inferred from the annual industrial share of total freshwater withdrawals and annual total freshwater withdrawals. We retain the most recent non-missing observation for each country to maximize spatial coverage. Technology-specific water-consumption intensities are then used to estimate additional water demand from onsite generation [48] and data-centre cooling.

### *2.1.5 Digital connectivity layer*

The digital connectivity layer represents whether candidate locations can receive transferred AI workloads through global fibre networks. It integrates terrestrial fibre networks, submarine cable systems, landing stations and internet exchange points. Terrestrial fibre networks are obtained from the International Telecommunication Union Interactive Transmission Map [49], while submarine cable systems and landing stations are derived from the Global Submarine Cable Map hosted on ArcGIS Online [50].

### *2.1.6 Land layer*

The land layer captures the economic and physical constraints imposed by land on data-centre siting and supporting energy infrastructure. It includes two components: land-availability constraints for renewable and other generation technologies and industrial land prices for data-centre construction.

1) land availability and renewable potential

Wind and solar deployment limits are based on the global renewable-energy potential dataset [51], which provides technology-specific installable capacity and generation potential. These estimates are used to constrain the maximum deployable wind and solar capacity in each region. For other generation technologies, available land is estimated at a 300 m×300 m grid-cell resolution using ESA WorldCover 2021 [52], combined with protected-area data from the World Database on Protected Areas [53]. We exclude land that is unsuitable or environmentally constrained, including protected areas, forests and other high-ecological-value ecosystems, wetlands and water bodies, steep terrain, highly urbanized or built-up areas and, under conservative assumptions, cropland or high-value agricultural land. The remaining suitable land area is converted into technology-specific upper bounds on deployable generation capacity.

2) Industrial land prices

Industrial land prices are compiled at the subnational region level for major countries using recent market reports,

commercial real-estate platforms, official investment documents and government publications. These sources provide regional observations that approximate prevailing industrial land market conditions. Owing to differences in national data availability, multiple data sources are integrated. In countries such as Australia, prices are primarily derived from structured market reports published by major real-estate service providers, whereas in parts of Europe and the United States, estimates are inferred from industrial land listings on national real-estate platforms. Where available, these estimates are cross-validated against benchmark land-value documents released by governments or investment agencies.

### *2.2. Construction cost quantification*

#### *2.2.1 Data-centre construction costs*

Construction costs of grid-supplied data centres are defined within a facility-side cost scope consistent with international data-centre cost benchmarks[54], [55]. The construction costs include general requirements and contractor fees, building-related costs, mechanical systems costs and electrical system costs. It excludes servers, networking hardware, AI accelerators, software, and tenant-owned IT equipment. Construction costs are expressed in US dollars per watt of IT capacity.

$$u_r^{\mathrm{GS}} = u_r^{\mathrm{shell}} + u_r^{\mathrm{mech}} + u_r^{\mathrm{elec}} + u_r^{\mathrm{gc}} \tag{1}$$

For each component, we first define a global mean unit cost and then apply component-specific regional adjustment factors. General requirements and general contractor fees are calculated as 10% of pre-fee construction costs [54]. Building-related costs are adjusted using the World Bank International Comparison Program construction price-level index [56], which captures cross-country differences in construction prices. Mechanical-system costs cover globally traded equipment, including chillers, pumps, fans, cooling towers and CRAH/CRAC units, and are adjusted using international import-price information from the World Integrated Trade Solution database [57]. Because the choice of space-conditioning technology substantially affects mechanical-system capital costs, accounting for 22-33% [54] of total construction costs, we develop a component-based cooling-cost model for the ten space-conditioning technologies described in Supplementary Text 1. This model combines equipment-level costs to estimate technology-specific capital-cost indices [58]-[60], which are then weighted by the technology mix determined by the siting optimization model. Water availability and local energy costs shape technology selection through climate-dependent PUE and WUE. Electrical system costs are adjusted from a global benchmark by incorporating grid reliability, which determines UPS backup and redundancy requirements.

Beyond these four components, the model incorporates two additional regional cost adjustments: (a) gaps in natural-gas network access are converted into the costs of pipeline connections and metering stations; and (b) digital connectivity conditions are represented through the capital costs of fibre interconnections and internet-exchange access. Together with the four base components, these adjustments define region-specific construction costs for grid-supplied configurations. For data centres with on-site power generation, construction costs additionally include land costs and the capital costs of constructing on-site power-generation facilities. Details are provided in Supplementary Text 2.

We compare the modelled construction costs with publicly available cost estimates for 51 major global data-centre markets, including Tokyo, Singapore, London, Frankfurt and Northern Virginia. As our results are reported at the subnational regional level, while existing market data are available primarily for city-level data-centre markets, the comparison provides an external consistency check rather than a strict one-to-one validation. Regional deviations range from 6.2% to 13.9%, with a global normalized deviation of 11.5% (see Supplementary Text 3).

*2.2.2 Financial feasibility screening*

Beyond physical resource constraints, the economic feasibility of data-centre deployment also depends on the cost of capital. Regions with otherwise favourable physical conditions may still be financially infeasible if financing costs are prohibitively high. To account for this factor, we screen candidate locations under two financing models: local financing and multinational financing. The local-financing model uses country-specific weighted average costs of capital (WACC) to reflect the financing terms faced by investments in each country. The multinational-financing model uses the cost of capital for international AI-related investment, adjusted by country risk premium. Locations are excluded from the siting framework if their capital costs exceed expected returns. Details of the screening procedure are provided in Supplementary Texts 6 and 7.

### *2.3. Computation-transfer-based siting model*

We develop a global computation-transfer-based siting model to identify the optimal siting of data-centre locations and capacity under energy, water, digital connectivity and land constraints. The model is formulated as an hourly optimization problem that links long-term planning with short-term operation by optimizing energy flows. It first embeds global data-centre construction costs, and then optimizes siting decisions by minimizing total costs and data-centre delay capacity.

*2.3.1 Objective function*

The objective function minimizes total system costs and delayed data-centre capacity under coupled energy, water, digital connectivity and land constraints.

$$minZ = C^{\mathrm{tot}} + \lambda^{\mathrm{DC}} \sum_{r} K_r^{\mathrm{DC,uns}} \tag{2}$$

Total annualized costs comprise annualized data-centre construction costs and annual operating costs across all locations, with 2030 used as the representative operating year. Construction costs are annualized using the region- and financing-specific cost of capital and the assumed economic lifetime.

$$C^{\mathrm{tot}} = \sum_{r} \left[\frac{\rho_{r,m}(1+\rho_{r,m})^N}{(1+\rho_{r,m})^N - 1} C_r^{\mathrm{DC}} + C_r^{\mathrm{opr}}\right] \tag{3}$$

$$C_r^{\mathrm{DC}} = u_r^{\mathrm{GS}} \cdot K_r^{\mathrm{DCgrid}} + u_r^{\mathrm{SG}} \cdot K_r^{\mathrm{DCself}} \tag{4}$$

$$C_r^{\mathrm{opr}} = \sum_{g} \left(c_{r,g}^{\mathrm{FOM}} \cdot K_{r,g}^{\mathrm{self,gen}} + \sum_{t} c_{r,g}^{\mathrm{VOM}} \cdot G_{r,g,t}^{\mathrm{self,gen}}\right) + \sum_{t} c_r^{\mathrm{grid}} \cdot G_{r,t}^{\mathrm{grid}} \tag{5}$$

$$\sum_{r} \left(K_r^{\mathrm{DC,uns}} + K_r^{\mathrm{DCgrid}} + K_r^{\mathrm{DCself}}\right) = \sum_{r} \left(K_r^{\mathrm{DC,demand}} - K_r^{\mathrm{DC}}\right) \tag{6}$$

For local financing, $\rho_{r,m}$ is given by the country-specific weighted average cost of capital [61]. For multinational financing, $\rho_{r,m}$ is defined as the international AI-capital discount rate [62] plus the country risk premium [63]. The annual operating costs include fixed operation and maintenance costs, variable operation and maintenance costs, and electricity costs.

*2.3.2 Energy layer constraints*

We model conventional electricity demand and data-centre workloads as flexible loads within daily operating cycles, but distinguish between their flexibility dimensions: the former can be shifted temporally across hours within each location, whereas the latter can be shifted both temporally and spatially across locations.

$$0 \le L_{r,t}^{\mathrm{up}}, L_{r,t}^{\mathrm{dn}} \le \alpha^{\mathrm{load}} L_{r,t} \tag{7}$$

$$\eta^{\mathrm{load}} \sum_{t \in T_d} L_{r,t}^{\mathrm{up}} = \sum_{t \in T_d} L_{r,t}^{\mathrm{dn}} \tag{8}$$

$$0 \le \begin{cases} D_{r,t}^{\mathrm{te,up}} + D_{r,t}^{\mathrm{sp,as,up}} \\ D_{r,t}^{\mathrm{te,dn}} + D_{r,t}^{\mathrm{sp,as,dn}} \end{cases} \le \phi_r^{\mathrm{DC}} \alpha^{\mathrm{DC}} (K_r^{\mathrm{DCgrid}} + K_r^{\mathrm{DCself}} + K_r^{\mathrm{DC}}) \tag{9}$$

$$0 \le D_{r,t}^{\mathrm{sp,up}}, D_{r,t}^{\mathrm{sp,dn}} \le \phi_r^{\mathrm{DC}} (1 - \alpha^{\mathrm{DC}} - 5\%)(K_r^{\mathrm{DCgrid}} + K_r^{\mathrm{DCself}} + K_r^{\mathrm{DC}}) \tag{10}$$

$$\eta^{\mathrm{te,DC}} \sum_{t \in T_d} D_{r,t}^{\mathrm{te,up}} = \sum_{t \in T_d} D_{r,t}^{\mathrm{te,dn}} \tag{11}$$

$$\eta^{\mathrm{sp,DC}} \sum_{j \in G_r} F_{rj,t}^{\mathrm{as}} = D_{r,t}^{\mathrm{sp,as,dn}} \tag{12}$$

$$\sum_{i \in G_r} F_{ir,t}^{\mathrm{as}} = D_{r,t}^{\mathrm{sp,as,up}} \tag{13}$$

$$\eta^{\mathrm{sp,DC}} \sum_{j \in N_r} F_{rj,t} = D_{r,t}^{\mathrm{sp,dn}} \tag{14}$$

$$\sum_{i \in N_r} F_{ir,t} = D_{r,t}^{\mathrm{sp,up}} \tag{15}$$

where $\eta^{\mathrm{load}}$ denotes the energy efficiency associated with temporal shifting of conventional electricity demand, and $T_d$ denotes the set of hourly time steps within day $d$, and $\eta^{\mathrm{te,DC}}$ and $\eta^{\mathrm{sp,DC}}$ are effective efficiencies accounting for the communication energy overheads of temporal and spatial workload shifting, respectively.

Following estimates from the U.S. Federal Energy Regulatory Commission and the IEA, $\alpha^{\mathrm{load}}$ is capped at 5% for conventional load [64], [65]. For data centres, we distinguish three categories of computational workloads according to their flexibility. First, fully flexible workloads can be shifted temporally or transferred globally without latency constraints. $\alpha^{\mathrm{DC}}$ is set to 10% for non-accelerated-server workloads and 25% for accelerated-server workloads [4], [66], with the latter serving as a proxy for AI-related workloads and accounting for 15% of total data-centre electricity demand. Second, partially flexible workloads can be transferred spatially only between locations satisfying latency requirements. Third, ultra-low-latency workloads, such as closed-loop robotic control, real-time machine vision and safety-critical autonomous systems, must remain locally or edge served. These workloads are assumed to account for 5% of computing demand [67], [68] and are excluded from spatial transfer.

In summary, the net load curves of data-centre workloads and conventional load can be expressed as

$$0 \le D_{r,t}^{\mathrm{net}} \le K_r^{\mathrm{DC}} + K_r^{\mathrm{DCgrid}} + K_r^{\mathrm{DCself}} \tag{16}$$

$$D_{r,t}^{\mathrm{net}} = D_{r,t}^{\mathrm{net,grid}} + D_{r,t}^{\mathrm{net,self}} = D_{r,t}^{\mathrm{DC}} + D_{r,t}^{\mathrm{te,up}} - D_{r,t}^{\mathrm{te,dn}} + D_{r,t}^{\mathrm{sp,up}} - D_{r,t}^{\mathrm{sp,dn}} + D_{r,t}^{\mathrm{sp,as,up}} - D_{r,t}^{\mathrm{sp,as,dn}} \tag{17}$$

$$L_{r,t}^{\mathrm{net}} = L_{r,t} + L_{r,t}^{\mathrm{up}} - L_{r,t}^{\mathrm{dn}} \tag{18}$$

Hourly generation is bounded by technology-specific installed capacity. For wind generation and solar generation, available power is further constrained by hourly resource conditions, represented by technology-specific capacity factors [69].

$$0 \le G_{r,g,t}^{\text{self,gen}} \le K_{r,g}^{\text{self,gen}} \tag{19}$$

$$0 \le G_{r,\text{wnd},t}^{\text{self,gen}} \le K_{r,\text{wnd}}^{\text{self,gen}} \phi_{r,t}^{\text{wnd}} \tag{20}$$

$$0 \le G_{r,\text{sol},t}^{\text{self,gen}} \le K_{r,\text{sol}}^{\text{self,gen}} \phi_{r,t}^{\text{sol}} \tag{21}$$

Power-system operation is modeled using the PyPSA-Earth framework [70], accounting for nodal power balance and transmission flows governed by voltage phase-angle differences and line reactance. Power flows are constrained by existing transmission capacity and planned expansion, with expansion potentials compiled from government plans and previous studies; Where such information is unavailable, transmission capacity is allowed to increase to at most three times its near-present level by 2050 [71], and use linear interpolation to derive the corresponding capacity by 2030.

$$-(F_l^{\text{ex}} + \Delta K_l^{\text{tr}}) \le F_{l,t} \le +(F_l^{\text{ex}} + \Delta K_l^{\text{tr}}) \tag{22}$$

Storage provides intertemporal balancing by charging during surplus periods and discharging when demand exceeds local supply. Its operation is constrained by installed capacity, charging and discharging limits, and state-of-charge dynamics.

$$0 \le P_{r,t}^{\text{ch,self/grid}}, P_{r,t}^{\text{dis,self/grid}} \le \eta^{\text{st}} K_r^{\text{st}} \tag{23}$$

$$S_{r,t} = S_{r,t-1} + \eta^{\text{ch}} P_{r,t-1}^{\text{ch,self/grid}} - P_{r,t-1}^{\text{dis,self/grid}} / \eta^{\text{dis}} \tag{24}$$

$$\underline{s} K_r^{\text{st}} \le S_{r,t} \le \overline{s} K_r^{\text{st}} \tag{25}$$

Hourly operation is governed by separate power-balance constraints for the public grid, grid-supplied data centres and self-generation data centres. Electricity purchases link grid-supplied data centres to the public grid, while onsite generation and storage supply self-generation data centres.

$$G_{r,t}^{\text{grid}} + P_{r,t}^{\text{dis,grid}} - P_{r,t}^{\text{ch,grid}} = \text{PUE}_r D_{r,t}^{\text{net,grid}} \tag{26}$$

$$\sum_g G_{r,g,t}^{\text{grid,gen}} - L_{r,t}^{\text{net}} - G_{r,t}^{\text{grid}} = \sum_l B_{r,l} F_{l,t} \tag{27}$$

$$\sum_g G_{r,g,t}^{\text{self,gen}} + P_{r,t}^{\text{dis,self}} - P_{r,t}^{\text{ch,self}} = \text{PUE}_r D_{r,t}^{\text{net,self}} \tag{28}$$

where $G_{r,g,t}^{\text{grid,gen}}$ denotes the power output of generation technology *g* at location *r* and time *t*.

### *2.3.3 Water layer constraints*

Data-centre water demand includes not only cooling water for data centres but also water use for behind-the-meter generation. Regional water demand is therefore defined as the sum of cooling water use, estimated from water use efficiency, and generation-related water use, estimated from onsite generation output or fuel consumption.

$$\sum_t (\text{WUE}_r D_{r,t}^{\text{net}} + \sum_g G_{r,g,t}^{\text{self,gen}} \omega_g) \le \alpha_r W_r^{\text{avail}} \tag{29}$$

*2.3.4 Digital connectivity layer constraints*

To ensure service quality, we impose latency constraints on all latency-sensitive workloads. Specifically, the end-to-end transmission delay between origin node *r* and destination node *j* must not exceed the service-specific threshold $L^{\max}$, set at 100 ms following International Telecommunication Union recommendations [72], [73]. Unlike dynamic workload migration between data centres, computation-transfer-based siting allows remotely located data centres to directly serve demand origins, avoiding additional inter-data-centre workload transfer and synchronization delays. We therefore represent each service interaction by a single request-response round trip, with the associated network latency and processing delay included.

$$N_r = \left\{ j \in G_r : lat \cdot \alpha \cdot d_{rj} + L_s^{proc} \leq L^{max} \right\} \tag{30}$$

where fibre round-trip latency is set to $0.010\ \mathrm{ms\,km^{-1}}$ [74], route-inflation factor is set to 1.2 [76], and $d_{rj}$ is the transmission distance between nodes $r$ and $j$. Service-specific processing delay is set to $5\ \mathrm{ms}$ [75]. Latency thresholds determine the feasible transmission range for computation transfer. We therefore validate the modelled maximum transmission distances against independent benchmark estimates [76], finding deviations below 0.6%.

*2.3.5 Land layer constraints*

For generation technologies, the deployable capacity is constrained by land availability, which can be calculated as:

$$A_r^{avail} = \sum_{p \in r} a_p \cdot M_p \tag{31}$$

where $M_p$ is a binary land-suitability mask that equals 1 for pixels suitable for generation deployment and 0 for excluded pixels, including protected areas, forests and other high-ecological-value ecosystems, wetlands and water bodies, steep terrain, highly urbanized or built-up areas and, under conservative assumptions, cropland and high-value agricultural land. Therefore, the land-constrained self-generation capacity in each location can be expressed as:

$$0 \leq K_{r,g}^{\mathrm{self,gen}} \leq \overline{K}_{r,g}^{\mathrm{self,gen}} \tag{32}$$

$$\sum_g A_{g,r}^{\mathrm{land}} K_{r,g}^{\mathrm{self,gen}} \leq A_r^{avail} \tag{33}$$

For wind and solar generation, the maximum installed capacity $K_{r,\mathrm{wnd}}^{\mathrm{self,gen}}$ and $K_{r,\mathrm{sol}}^{\mathrm{self,gen}}$ are constrained not only by land availability but also by resource quality. We therefore use the renewable-energy potential dataset developed by the German Aerospace Center [51] to further bound the maximum installable capacities of wind and solar power.

$$K_{r,\mathrm{wnd}}^{\mathrm{self,gen}} \leq \overline{K}_{r,\mathrm{wnd}}^{\mathrm{self,gen}} \tag{34}$$

$$K_{r,\mathrm{sol}}^{\mathrm{self,gen}} \leq \overline{K}_{r,\mathrm{sol}}^{\mathrm{self,gen}} \tag{35}$$

## 3. Results

We construct 36 scenarios by combining four data-centre development pathways with nine climate-policy pathways. Each combined scenario is labelled using the format DxCy, where Dx denotes the data-centre development pathway and Cy denotes the climate-policy pathway. For example, D1C3 represents the Headwinds Case under the 1.5 °C immediate mitigation pathway. The data-centre pathways are based on the IEA Energy and AI scenarios and span a wide range of future AI and digital-service demand, from the low-growth Headwinds Case to the high-growth Lift-Off Case, with the High Efficiency and Base cases representing efficiency-driven and central trajectories, respectively. The climate-policy pathways are taken from the ADVANCE scenario framework and cover immediate mitigation, delayed mitigation, nationally determined contributions, no-policy and reference pathways, as shown in Table 1.

**Table. 1 Scenario design used in this study**

| Scenario group | Code | Scenario name | Description |
|---|---|---|---|
| Data-centre development | D1 | Headwinds | Slow growth in AI and digital-service demand, weaker investment, and stronger local constraints. |
| Data-centre development | D2 | High Efficiency | Digital-service demand follows the Base Case, but energy demand is reduced through lower PUE, software optimisation, right-sized models and more efficient specialised hardware. |
| Data-centre development | D3 | Base | Central pathway reflecting continued growth in AI and digital services, data-centre deployment and efficiency improvements under current expectations. |
| Data-centre development | D4 | Lift-Off | Stronger AI adoption and digital-service demand drive faster data-centre deployment. Greater siting flexibility, adaptive supply chains and increased onsite generation reduce local constraints. |
| Climate policy | C1 | Med2C-2020 | Immediate mitigation pathway towards a medium-stringency 2 °C target. |
| Climate policy | C2 | WB2C-2020 | Immediate mitigation pathway towards a well-below 2 °C target. |
| Climate policy | C3 | 1.5C-2020 | Immediate mitigation pathway towards a 1.5 °C target by 2100. |
| Climate policy | C4 | Med2C-2030 | Delayed mitigation towards a medium-stringency 2 °C target. |
| Climate policy | C5 | WB2C-2030 | Delayed mitigation towards a well-below 2 °C target. |
| Climate policy | C6 | Price1.5C-2030 | Delayed mitigation with a carbon-price pathway consistent with a 1.5 °C target. |
| Climate policy | C7 | INDC | Policy pathway based on intended nationally determined contributions. |
| Climate policy | C8 | NoPolicy | No additional climate-policy pathway. |
| Climate policy | C9 | Reference | Reference pathway representing the continuation of existing policies. |

### *3.1. Quantification of data-centre construction costs*

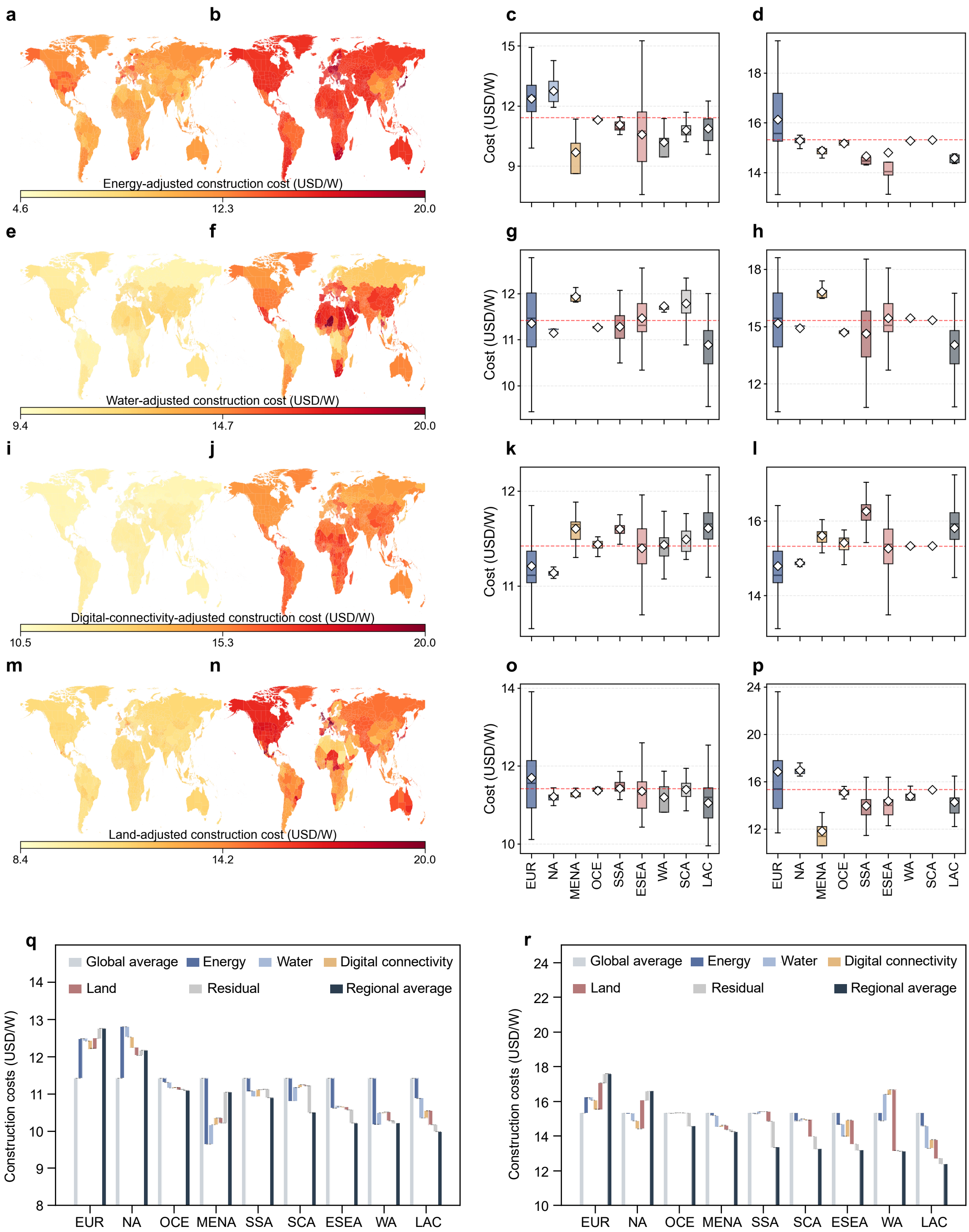

**Fig. 2 Data-centre construction costs under two deployment modes. a-d**, Energy-adjusted construction costs: global maps of construction costs under grid-supplied (a) and self-generation (b) modes, and regional distributions of construction costs under grid-supplied (c) and self-generation (d) modes in nine study regions. Construction costs [54], [55] are region-specific costs per watt of IT capacity, including facility-side construction, mechanical and electrical systems, land and resource-related cost adjustments, but excluding servers, networking hardware and AI accelerators; details are provided in Supplementary Text 2. **e-h**, Water-adjusted construction costs, presented in the same four-panel structure. **i-l**, Digital-connectivity-adjusted construction costs. **m-p**, Land-adjusted construction costs. Box plots show the median (horizontal line), and interquartile range across subnational regions within each region; open diamonds indicate the regional mean, and red dashed horizontal lines indicate the global mean. **q**, Unit construction cost of grid-supplied data centres, after accounting for spatial heterogeneity in energy, digital connectivity, water and land factors. Cost increment is decomposed into contributions these four factors. **r**, Unit construction cost of self-generation data centres across nine study regions.

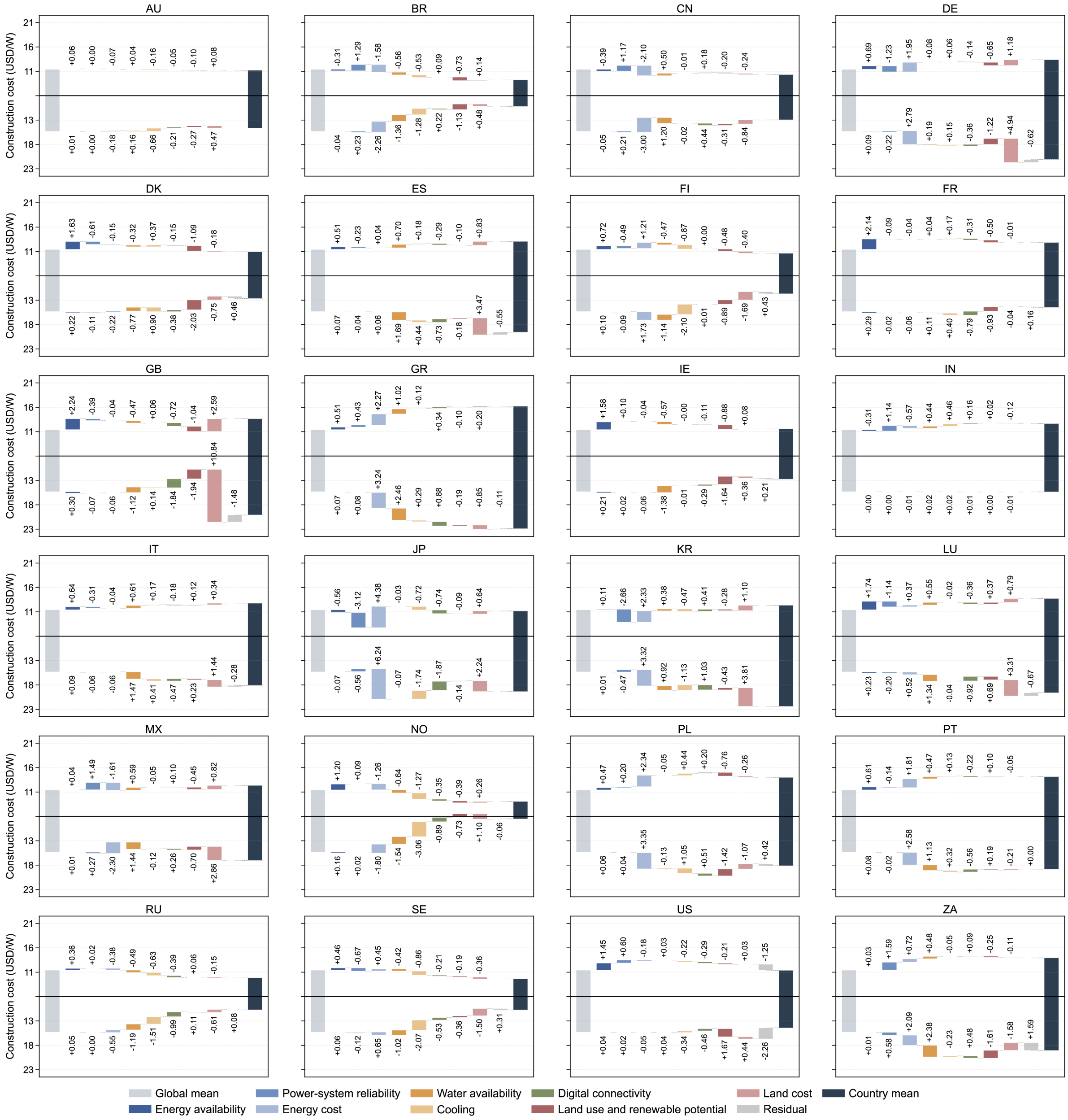

**Fig. 3 Regional data-centre construction costs under grid-supplied and self-generation deployment.** Waterfall decomposition of average unit construction costs for major countries under grid-supplied deployment (upper bars) and self-generation deployment (lower bars). Light-grey bars denote the global average construction costs; colored bars show additive country-specific deviations attributable to energy availability, power-system reliability, energy cost, water availability, cooling, digital connectivity, land use and renewable potential, land cost and residual terms.

We first provide the global quantification of data-centre construction costs across more than 4,000 subnational regions worldwide (Fig. 2), accounting for energy, digital connectivity, water and land factors, as well as the optimal choice of cooling system. Costs are computed under two deployment modes: grid-supplied [79], where data centres are powered by power systems, and self-generation [80], which uses dedicated on-site or nearby supply and is increasingly adopted to avoid connection delay [80]. To benchmark the cost estimates, we compare modelled regional construction costs with publicly available estimates for 51 major data-centre markets, including Tokyo, London, Frankfurt and Northern Virginia. Because our model reports costs at the subnational regional scale whereas market data are typically available for city-level data-centre markets, this comparison serves as an external consistency assessment. The resulting regional deviations range from 6.2% to 13.9%, with a global normalized deviation of 11.5% (Supplementary Text 3).

Under grid-supplied mode, energy-related factors dominate regional differences in construction costs. In Europe and North America, they add approximately 1.06-1.38 USD $W^{-1}$ to unit construction costs, but through different mechanisms. In Europe, the increase is split between grid limitation and energy costs, each contributing about 0.50 USD $W^{-1}$. In North America, the increase is driven mainly by grid limitation, contributing up to 1.13 USD $W^{-1}$, with a further 0.43 USD $W^{-1}$ associated with power-system reliability as weather-related disruptions increase the need for backup capacity. Limited digital connectivity imposes a modest but spatially concentrated cost increment, primarily in Western Asia and sub-Saharan Africa, where fibre-constrained regions require new access links with associated cost increment reaching about 0.18 USD $W^{-1}$ (Fig. 2i,k,q). Land and water factors contribute somewhat larger costs in some regions, but remain secondary relative to the energy layers, with contributions below 0.52 USD $W^{-1}$ even in North Africa and Latin America (Fig. 2f,h,n,p,q,r).

Under self-generation mode, land-related factors become the key determinant of construction costs because behind-the-meter power supply is internalized into the siting decision. The land price is most pronounced in Europe, where land-price-related factors add up to 1.81 USD $W^{-1}$ to unit construction cost. Although favorable renewable-resource potential reduces costs by 0.30 USD $W^{-1}$, this benefit is insufficient to offset the large land-price-related cost increment, leaving a net land cost increment of around 1.51 USD $W^{-1}$ (Fig. 2r). In North America, the high land-related cost increment is driven primarily by the United States, reaching 1.63 USD $W^{-1}$ (Fig. 2r) at the regional level and 2.11 USD $W^{-1}$ (Fig. 3) in the United States. Because self-generation relies on dedicated behind-the-meter power supply, grid-related power factors play only a minor role in this mode (Fig. 2r). Water factors become more important under self-generation than under grid-supplied deployment (Fig. 2q,r). This is because water demand arises not only from data-centre cooling, but also from associated power generation. In water-stressed regions such as West Asia, water-related constraints increase unit construction cost by up to 1.50 USD $W^{-1}$ (Fig. 2r), highlighting that self-generation reduces grid dependence but can amplify exposure to land and water limitations.

### *3.2. Cost increases and residual connection delays under local energy expansion*

In locations affected by grid-connection delays, bringing new data-centre capacity online typically requires expanding transmission networks or deploying self-generation data centres to bypass connection constraints. Such local energy expansion has long been the primary strategy for relieving data-centre connection delays. We test this strategy across alternative data-centre development scenarios and climate-policy scenarios (Fig. 4). The results show that relying on local energy expansion to resolve data-centre connection delays entails substantial incremental costs, yet remains incomplete: by 2030, 10-15% of global data-centre capacity still faces connection delays even after local energy-system expansion (Appendix Fig. 1).

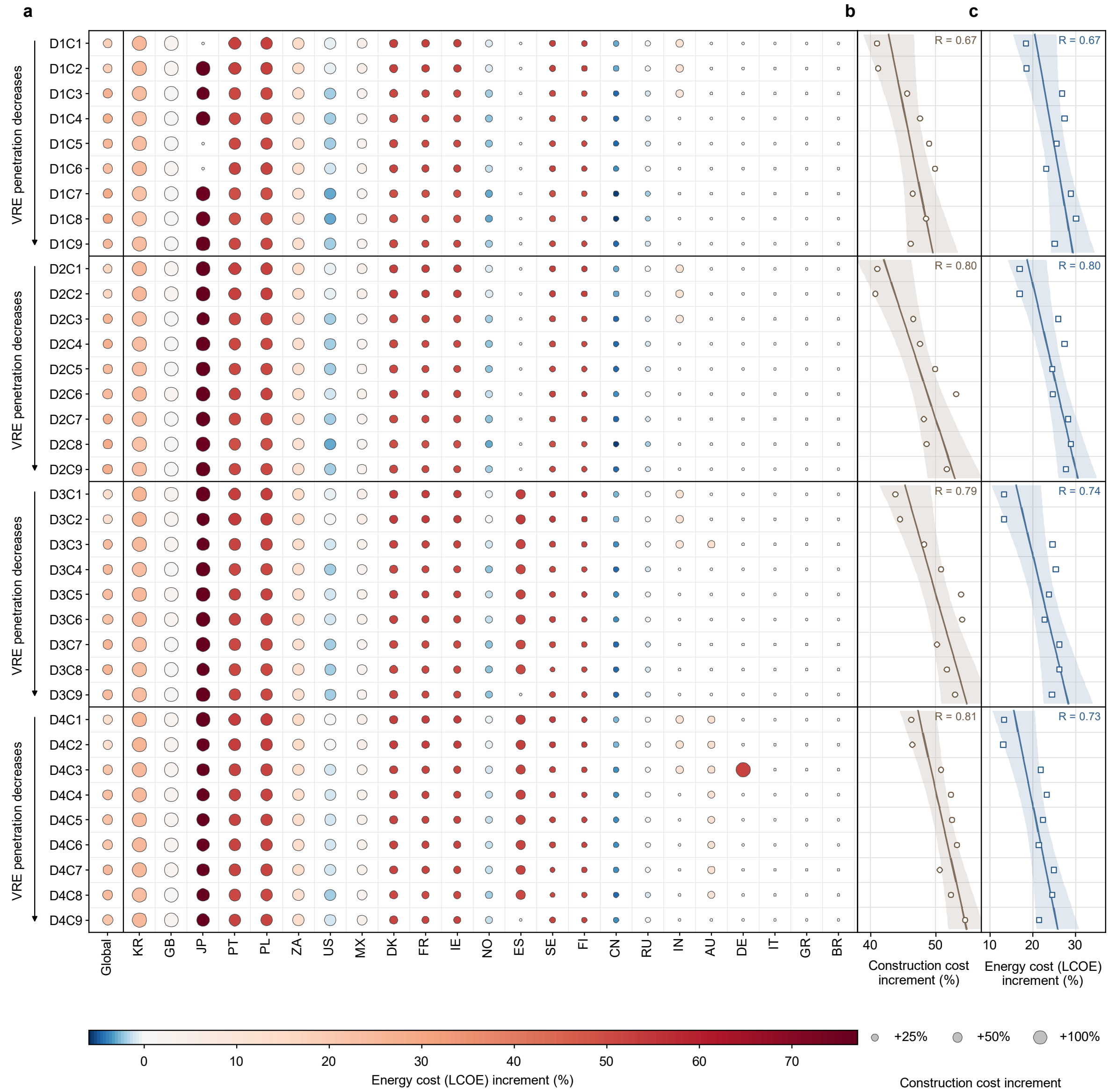

**Fig. 4 Cost increment of enabling data-centre deployment through local energy expansion. a,** Construction and energy cost increments across major countries under alternative data-centre development scenarios and climate-policy scenarios, shown for regions in the top 50% of connection-delay shares; scenario definition as shown in Table 1. Marker colour denotes the energy-cost (LCOE) increment, whereas marker size denotes the construction-cost increment. Within each case, scenarios are ordered by increasing variable renewable energy (VRE) penetration. **b,c**, Global cost increments across scenarios. Each point represents one scenario. Linear fits summarize the relationship between VRE penetration and the global b, construction-cost increment and c, energy-cost increment.

From a cost perspective, local energy expansion imposes a substantial construction-cost increment in delayed regions. Across all scenarios, regions in the top 50% of connection-delay shares exhibit data-centre construction costs that are 40% and 54% higher than other regions, respectively (Fig. 4a,b). We further assess the energy-cost implications of these delays using levelized costs of electricity (Fig. 4a,c), finding that global data-centre energy costs increase by 13-33% (Fig. 4c). However, there are important exceptions. In countries such as China, the United States and Norway, where energy resources are more readily accessible, delayed regions can exhibit lower energy costs. Higher future data-centre demand further amplifies construction costs (Fig. 4b): compared with D1-based scenario, D4 increases construction costs by around 5% globally, primarily because of greater local generation deployment (Fig. 4b). However, this additional local self-generation lowers data-centre energy costs, leading to an approximately 4% reduction in energy costs from D1 to D4 (Fig. 4c). Finally, we observe an approximately linear relationship between variable renewable energy (VRE) penetration and cost increments. In high-VRE scenarios, larger shares of wind and solar generation increase the demand for spatial balancing, transmission capacity and system flexibility [81], constraining effective grid capacity. As connection delays become more geographically widespread (Appendix Fig. 2), the associated construction-cost increments tend to converge towards the system-wide average. Europe deviates from this pattern because, across the climate-policy scenarios, nuclear generation expands faster than VRE generation [35], adding firm low-carbon capacity that reduces connection delays (Appendix Fig. 2).

### *3.3. Data-centre redistribution through computation-transfer siting*

Given the high cost and residual delays associated with local energy expansion, we next examine an alternative strategy: shifting computation rather than expanding energy supply around existing computing hubs. The computation-transfer-based siting framework reallocates future data-centre capacity from energy-constrained demand centres to more dispersed, resource-abundant locations while transferring computational workloads through digital networks. This strategy aims to bypass local grid-connection constraints by moving computation towards available energy resources.

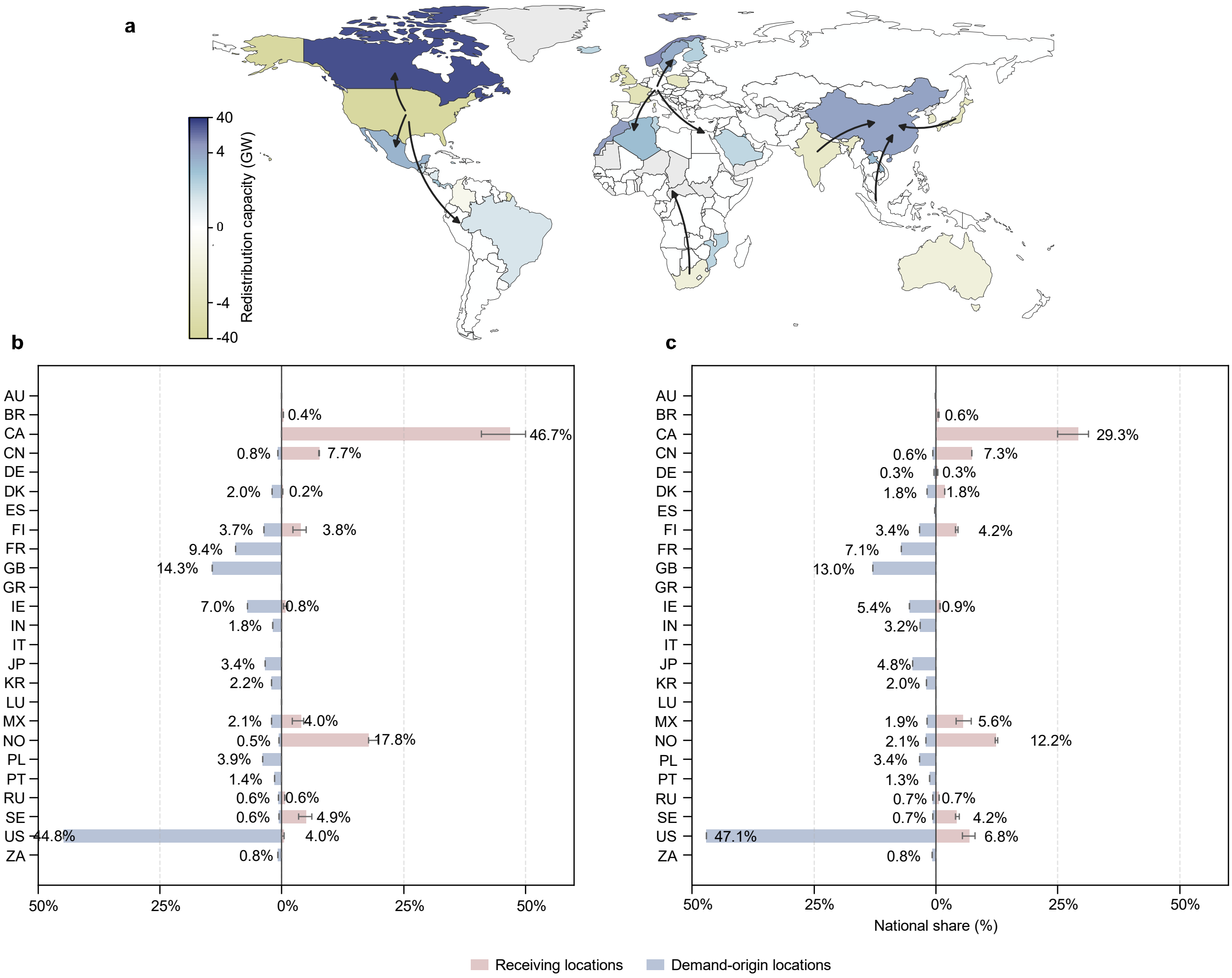

**Fig. 5 Data-centre redistribution under computation-transfer-based siting framework. a,** Global redistribution of data-centre capacity across the Headwinds (D1) and Lift-Off (D4) scenarios. Negative values indicate capacity relocated away from demand-origin locations; positive values denote the receiving locations. **b,c,** Regional shares of transferred data-centre capacity among major countries under the **b**, Headwinds and **c**, Lift-Off scenarios, expressed as percentages of total globally redistributed capacity. Error bars indicate the 5th–95th percentile range across climate-policy scenarios.

First, to ensure financial feasibility, we exclude locations with high cost-recovery risk under different financing modes, including local and multinational financing (Supplementary Text 6,7), mainly in parts of North Africa, southern Latin America, West Asia and Eastern Europe. Building on this, we quantify data-centre siting under global computation transfer. Nearly 90% (Appendix Fig. 3) of relocated capacity originates from established hubs in the United States and Europe, whereas Canada, Nordic countries, the Arabian Peninsula and Latin America collectively host most relocated data-centre capacity (Appendix Fig. 3). In the Headwinds scenario, the United States is the largest source of relocated capacity, accounting for 44.8% (Fig. 5b), followed by the United Kingdom (14.3%) and France (9.4%) (Fig. 5b). In the Lift-Off scenario, the share originating from the United States increases to 47.1% (Fig. 5c). Canada hosts 46.7% (Fig. 5b) of relocated capacity in the Headwinds scenario, supported by hydropower resources [77], cold-climate cooling potential [78] and low land costs. Under the Lift-Off scenario, however, its share falls to 29.3% (Fig. 5c), as higher global computation demand tightens the availability of suitable receiving locations. Europe shows a similar pattern. In the Headwinds scenario, Nordic countries host more than 60% of Europe's relocated capacity, led by Norway (17.8%), Sweden (4.9%) and Finland (3.8%) (Fig. 5b). Under the Lift-Off scenario, constraints on suitable receiving capacity in Nordic countries shift much of Europe's remaining relocated capacity to the Middle East (Fig. 5a), which hosts over 40% of the regional total.

Deployment modes also differ markedly across receiving regions. In Nordic countries, self-generation model consistently accounts for more than 60% of new capacity (Appendix Fig. 4). This reflects abundant low-cost renewable energy resources, together with competitive land costs, which make Nordic countries more attractive than Western and Central Europe for accommodating relocated self-generation data-centre capacity. By contrast, China accommodates almost all redistributed capacity through grid-supply model, with the share approximately 90% (Appendix Fig. 4). This pattern reflects recent expansion of transmission and generation infrastructure, which leaves sufficient headroom for utilities to integrate migrated data-centre capacity into the power system.

### *3.4. Costs and emissions benefits through computation-transfer siting*

Having identified the global redistribution pattern of data-centre capacity, we next quantify its implications for infrastructure investment, energy costs and carbon emissions. By leveraging global data transfer, new data centres can be built in more suitable regions, substantially reducing the energy-infrastructure investment required to integrate new data-centre capacity relative to local energy expansion. We compare computation-transfer-based siting with two representative local energy-expansion strategies: local self-generation to bypass connection delay, and transmission expansion around existing demand hubs to accommodate additional data-centre load.

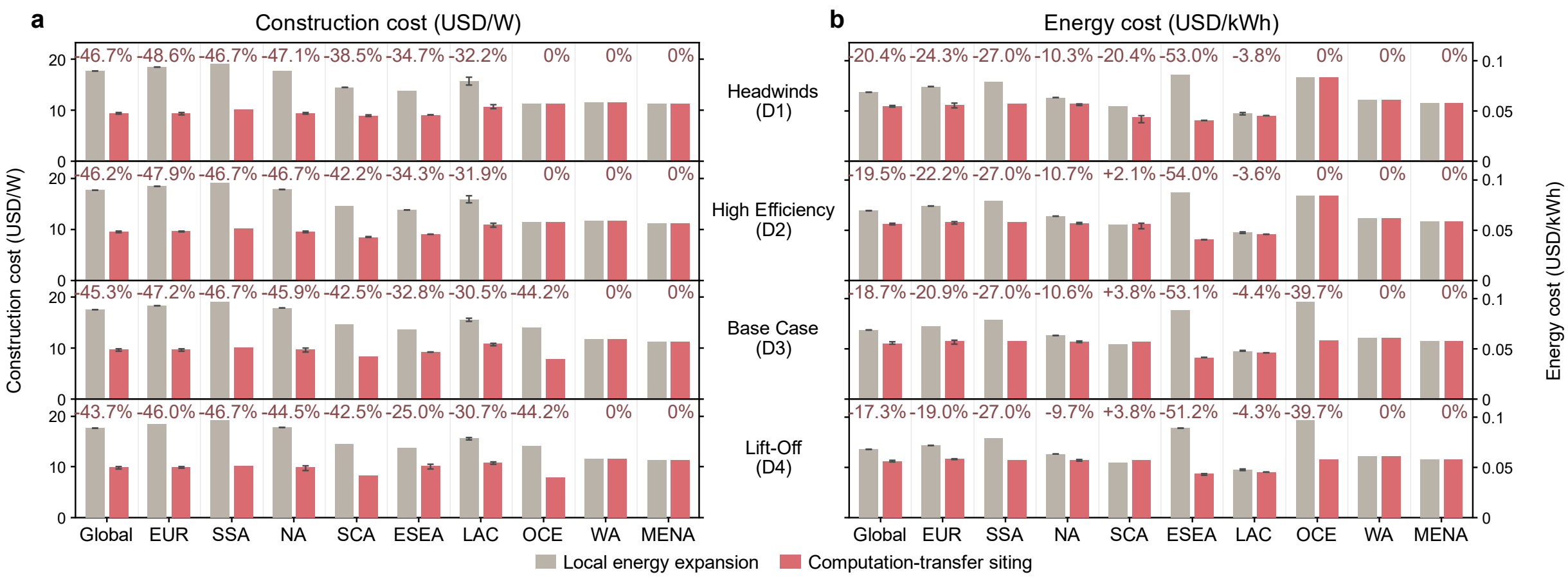

**Fig. 6 Regional reductions in data-centre construction and energy costs.** Cost reductions across nine regions under different data-centre development scenarios. **a**, Reduction in data-centre construction costs. **b**, Reduction in energy costs. Grey denotes energy-expansion strategies,

whereas red denotes computation-transfer-based siting strategies. Numbers in the upper of each cell indicate the percentage of cost reduction. Error bars indicate the 5th-95th percentile range across climate-policy scenarios.

Compared with local self-generation, computation-transfer-based siting markedly lowers data-centre construction costs. In the Headwinds scenario, global construction costs fall from 17.7 to 9.6 USD $W^{-1}$, a reduction of 45.8% (Fig. 6a). These savings arise from two mechanisms: part of the data-centre capacity is relocated to sites with available grid access, avoiding local generation equipment and associated land costs; and part is reallocated to sites with lower land and generation costs. Cost reductions are largest in Europe and NA, where connection pressure is high, reaching 48.6% and 47.1%, respectively (Fig. 6a). In the Lift-Off scenario, stronger global data-centre expansion tightens the availability of suitable regions, but computation-transfer-based siting still reduces construction costs by 43.7% (Fig. 6a). Beyond construction costs, computation-transfer-based siting reduces levelized costs of electricity by 17.3-20.4% (Fig. 6b), showing that it can both substitute for local generation investment and lower long-term operating energy costs by better matching computational demand with low-cost energy supply. Compared with transmission expansion, computation-transfer-based siting avoids the need for about 34-67 GW of additional transmission capacity (Appendix Fig. 5). More than 85% of this avoided transmission demand is concentrated in Europe and North America. Accounting for *N*-1 and *N*-2 transmission redundancy further increases the additional capacity required under conventional transmission expansion, with increments ranging from 5.2% in North America to 27.8% in sub-Saharan Africa. The avoided transmission capacity is comparable in magnitude to Italy's national peak electricity demand.

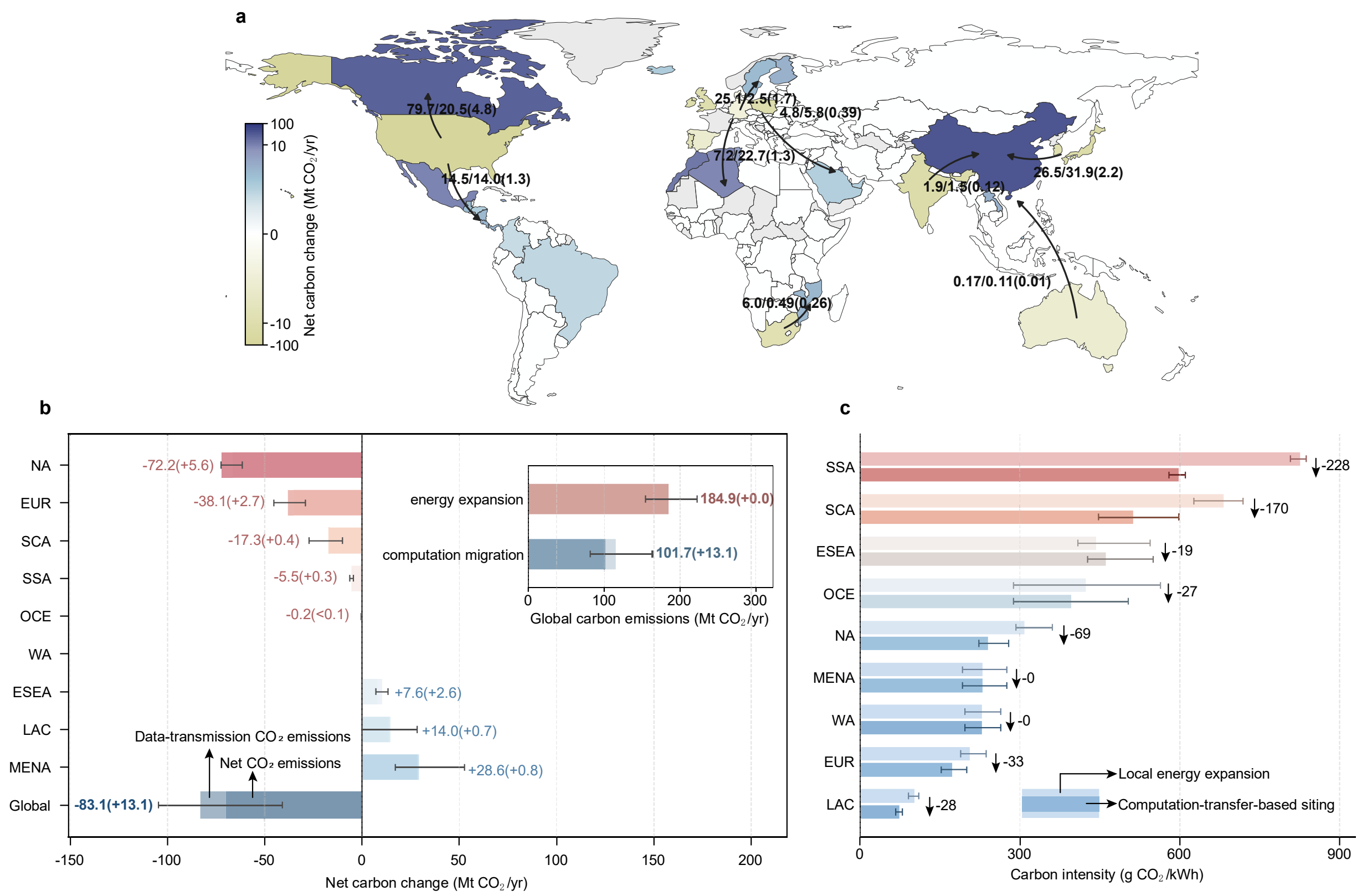

**Fig. 7 Environmental implications of computation-transfer-based siting. a,** Country-level changes in net annual $CO_2$ emissions caused by the redistribution of data-centre capacity under the computation-transfer-based siting. Arrows indicate major cross-border capacity redistribution. Labels beside each arrow report annual emissions in the format X/Y(Z). X denotes emissions under the local energy expansion, Y represents emissions under the computation-transfer-based siting, and Z denotes data-transmission emissions along that corridor, all in Mt $CO_2$ $yr^{-1}$. Thus, X-Y-Z gives the net $CO_2$ emissions effect of computation-transfer-based siting relative to local energy expansion. **b**, Regional net $CO_2$ emission changes under the computation-transfer-based siting compared with local energy expansion. The inset compares total $CO_2$ emissions from new data-centre capacity. **c**, Carbon intensity of new data centres. Light and dark colours represent local energy expansion and computation-transfer-

based siting, respectively. Numbers indicate the differences between the two strategies. Error bars indicate the 5th–95th percentile range across data-centre development and climate-policy scenarios.

Computation-transfer-based siting substantially reduces $CO_2$ emissions from new data-centre capacity. Relative to the local-energy expansion, Computation-transfer-based siting reduces annual $CO_2$ emissions from 184.9 to 114.8 (101.7+13.1) Mt $CO_2$ $yr^{-1}$ in 2030, a reduction of 37.9% (Fig. 7b). This reduction is concentrated mainly in North America (−72.2 Mt $CO_2$ $yr^{-1}$), driven by the relocation of US capacity to Canada, where grid emission factor is more than 60% lower than in the United States [82], [83]. However, this relocation also introduces additional emissions from data transmission, amounting to 4.8 Mt $CO_2$ $yr^{-1}$. Moreover, the global net reduction is accompanied by higher $CO_2$ emissions in receiving regions, including East Asia, Southeast Asia, Latin America and the Middle East. The Middle East and North Africa face the largest increase, with additional $CO_2$ emissions reaching 29.4 Mt $CO_2$ $yr^{-1}$, including 0.8 Mt $CO_2$ $yr^{-1}$ from data transmission (Fig. 7b).

The emissions benefits of computation-transfer-based siting are not universal. They depend on the interaction between grid carbon intensity and the optimal generation mix under the behind-meter generation of data centres. Carbon emission reductions are concentrated where data-centre capacity is reallocated from higher-carbon to lower-carbon grids, including from the United States to Canada (-54.4 Mt $CO_2$ $yr^{-1}$), from Central and Western Europe to the Nordic countries (-20.9 Mt $CO_2$ $yr^{-1}$) (Fig. 7a). By contrast, some computation transfer corridors generate higher emissions than they would under a scenario without computation-transfer siting. Capacity relocation from Europe to the Middle East increases emissions by 1.39 Mt $CO_2$ $yr^{-1}$, while relocation from Europe to North Africa increases emissions by 16.8 Mt $CO_2$ $yr^{-1}$. These reverse-gradient corridors emerge when cost optimization is driven primarily by non-carbon resource layers, such as land availability, water constraints or digital connectivity. The same mechanism explains the 29.4 Mt $CO_2$ $yr^{-1}$ increase in the Middle East and North Africa: the Arabian Peninsula is an economically attractive receiving region because of its energy [84] and connectivity advantages [85], but its current grid carbon intensity remains substantially higher than that of the European regions it serves.

**Uncertainty analysis**

We conduct a range of sensitivity analyses across computational demand, economic parameters, policy-implementation settings and climate-policy scenarios. We focus on three results: the effectiveness of computation-transfer-based siting in alleviating connection-delayed data-centre capacity, the geographic pattern of optimized siting and the magnitude of connection-delayed capacity (Supplementary Text 4). First, computation-transfer-based siting consistently alleviates connection-delayed data-centre capacity. The geographic allocation of optimized siting also remains broadly stable across economic, policy-implementation and climate-policy assumptions: source regions remain concentrated in the United States and Europe, while receiving regions vary only modestly with workload flexibility. In contrast, the magnitude of connection-delayed capacity is highly sensitive to climate-policy scenarios. More stringent climate policies accelerate VRE penetration, which increases the need for spatial balancing, transmission capacity and system flexibility, thereby constraining effective grid-connection capacity and increasing delayed capacity. This sensitivity is uneven across countries: France, and Norway show relatively stable exposure, whereas India, South Africa and Japan are much more sensitive to climate-policy pathways.

## 4. Discussion and Policy Implications

The rapid expansion of data centres is increasingly constrained by energy-supply bottlenecks, imposing substantial local cost burdens on data-centre construction and increasing the risk of connection delays. Our subnational-resolution analysis reveals that in regions facing connection-delayed capacity, data-centre construction costs are 40-54% higher and electricity costs 13-33% higher than regions without connection delays. We further find that energy-system strategies cannot fully resolve these delays: 10-15% of global data-centre capacity still faces connection delays by 2030. Here, by treating computation as transferable across regions, our analysis offers a new pathway for AI infrastructure deployment, in which new capacity is matched with regions of stronger resource endowments rather than tethered to demand hubs, with AI workloads routed accordingly. We evaluated this computation-transfer-based siting across physical, economic and environmental dimensions, and reached the following conclusions.

Physically, computation-transfer-based siting can effectively alleviate AI energy bottlenecks. More than 95% of the delayed capacity can be brought online by 2030. Economically, the locations facing the most severe connection delays are also those where local energy-system expansion is most costly. computation-transfer-based siting breaks this spatial coupling, reducing data-centre construction costs by 43.7-46.7% and energy costs by 17.3-20.4%. Environmentally, it lowers $CO_2$ emissions from new data-centre capacity by 37.9%. However, computation-transfer-based siting does not always follow low-carbon grid corridors, as siting also depends on energy, land, water and digital connectivity. For example, relocating data-centre capacity from Europe to North Africa increases emissions by 16.8 Mt $CO_2$ $yr^{-1}$.

Several boundary conditions merit further discussion. First, the effectiveness of computation-transfer-based siting in accelerating data-centre deployment depends on the share of latency-tolerant workloads and the latency thresholds of digital services. Sensitivity tests show that a 10% change in the share of latency-tolerant workloads changes the delayed capacity by less than 1.8%. Tightening the service-latency threshold from 100 ms to 50 ms increases the share of remaining delayed data-centre capacity by only around 3%. Second, we do not impose explicit capacity constraints on fibre networks, as currently lit submarine-cable capacity accounts for only about half of deployed design capacity [21]. Although aggregate submarine cable capacity remains sufficient in most cases, corridor-level analysis (Supplementary Text 8) shows that a small number of corridors, such as the Americas Region Caribbean Optical-Ring System (ARCOS) [86] could still face capacity constraints under a bandwidth-intensive multimodal transfer scenario. Third, our analysis focuses primarily on techno-economic factors and does not explicitly account for energy security, data security and environmental compensation issues associated with cross-border data-centre deployment.

These findings suggest three policy directions for next-generation AI infrastructure planning. First, planners should distinguish latency-tolerant workloads from latency-sensitive real-time inference and interactive services, thereby identifying which capacity must remain near demand hubs and which can be relocated to better-resourced regions. Second, global computation-transfer-based siting could relieve more than 95% of connection-delay capacity and substantially reduce energy-infrastructure investment. However, realizing this potential depends on removing national barriers to data flows and cross-border data-centre deployment, while ensuring coordinated governance of data security. Policy efforts should therefore strengthen coordination of cross-border regulations for data, computation and infrastructure. Finally, because computation-transfer-based siting redistributes environmental burdens across regions, AI infrastructure governance should combine digital networks and energy-system planning with frameworks for carbon accounting, carbon trading and environmental responsibility, so that global efficiency gains are not achieved at the expense of unaccounted environmental costs in receiving regions.

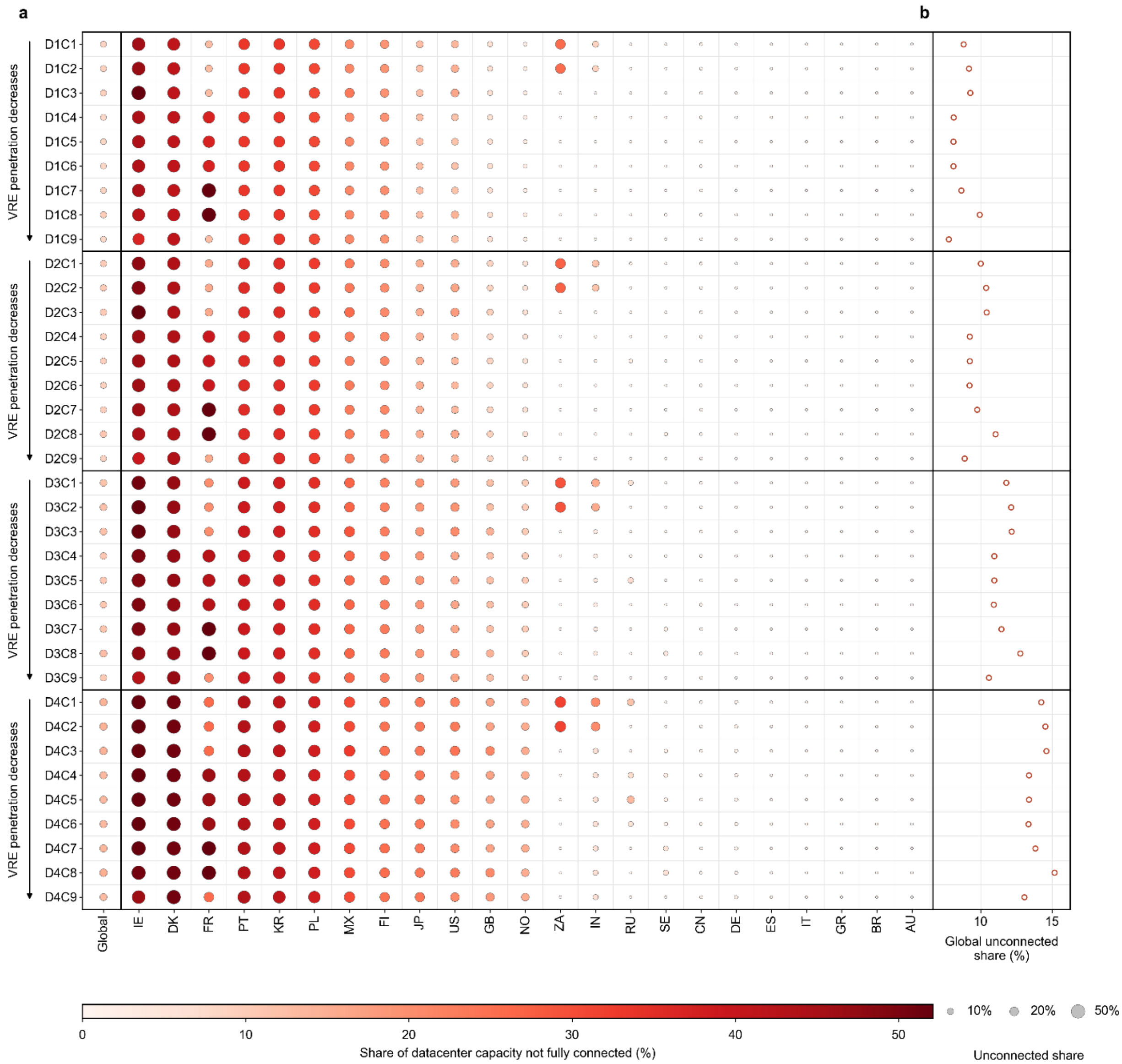

**Appendix Fig. 1 Residual unconnected data-centre capacity under local energy expansion. a,** Share of data-centre capacity that remains unconnected after local energy expansion across global and major country. Results are shown for combinations of data-centre development scenarios (D1-D4) and climate-policy scenarios (C1-C9), with C1-C9 ordered from higher to lower VRE penetration. Circle color and size indicate the unconnected share of data-centre capacity. **b**, Global unconnected share under the same scenario combinations.

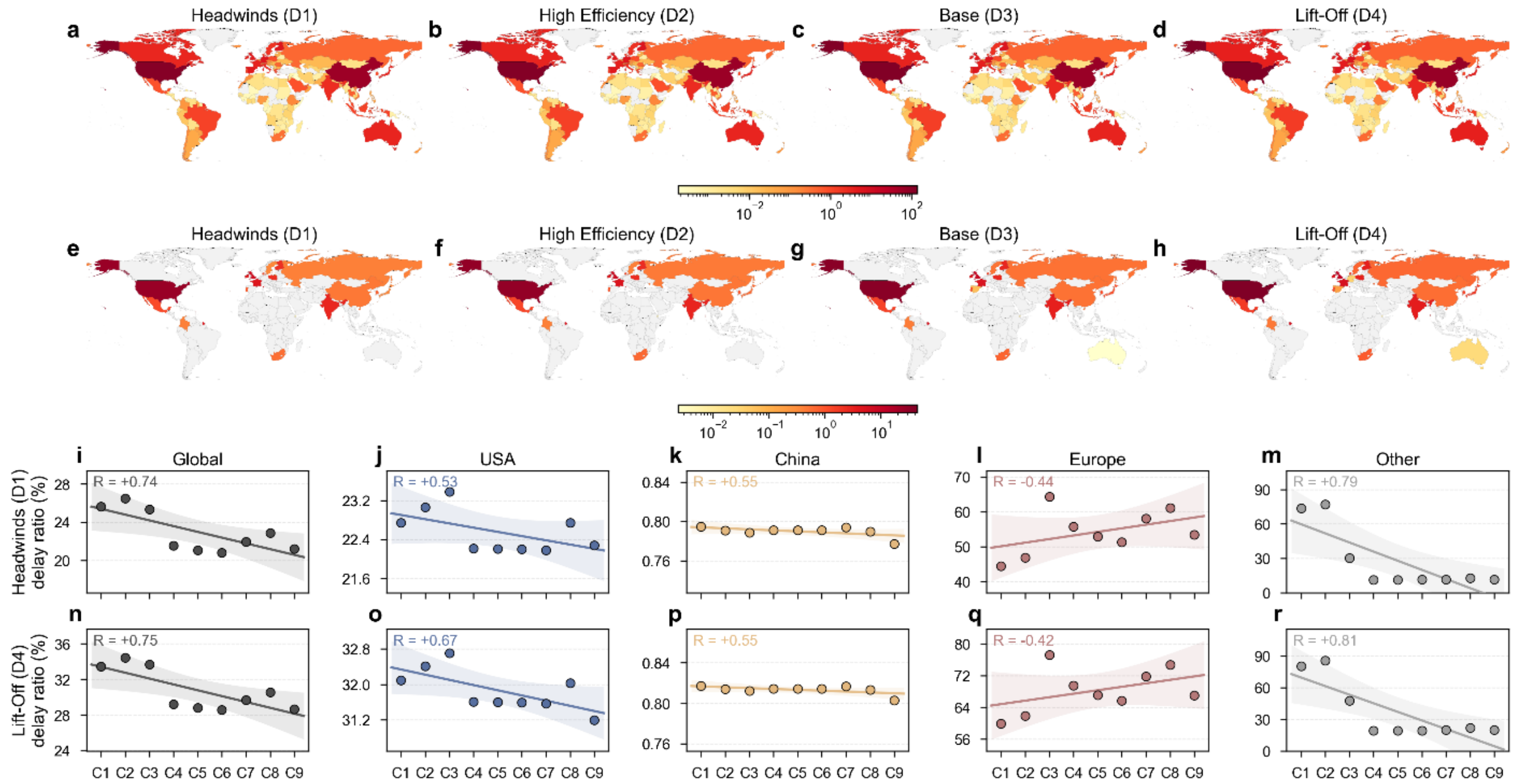

**Appendix Fig. 2 Global connection delay for data centres**. **a-d**, Projected data-centre capacity (GW) by country in 2030 under the Headwinds (D1), High Efficiency (D2), Base Case (D3) and Lift-Off (D4) scenarios, respectively. **e-h**, Projected connection-delayed data-centre capacity (GW) by country in 2030 under different scenarios. **i-m,** Connection-delay ratios under the Headwinds scenario for the global system, USA, China, Europe and other regions. **n-r**, Corresponding connection-delay ratios under the Lift-Off scenario. Climate-policy scenarios C1-C9 are ordered from higher to lower VRE penetration. Points denote individual scenarios, fitted lines show the trend across VRE penetration levels and shaded bands indicate uncertainty intervals.

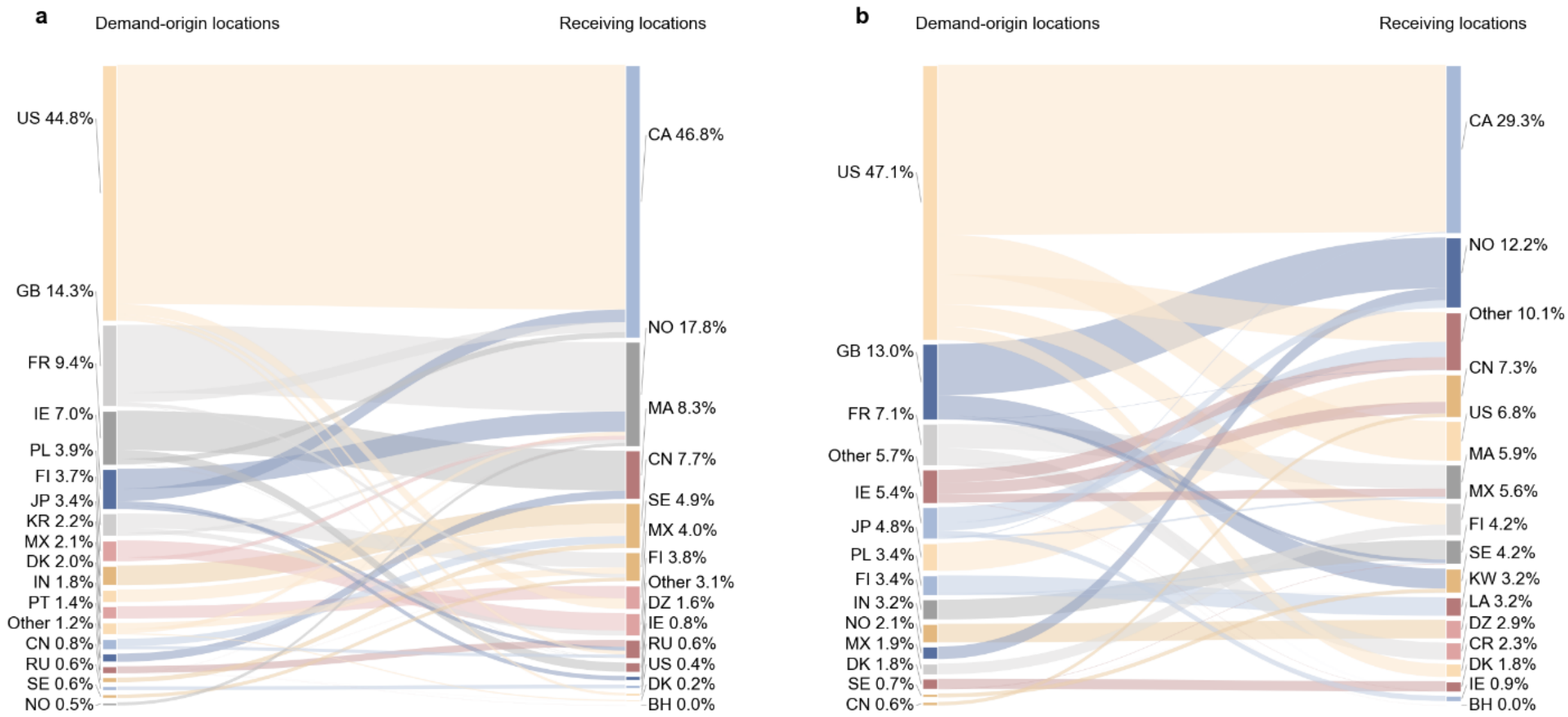

**Appendix Fig. 3 Data-centre capacity-redistribution shares under two development scenarios**. **a**, **b**, Sankey diagrams of country-level redistribution for the a, Headwinds and b, Lift-Off scenarios.

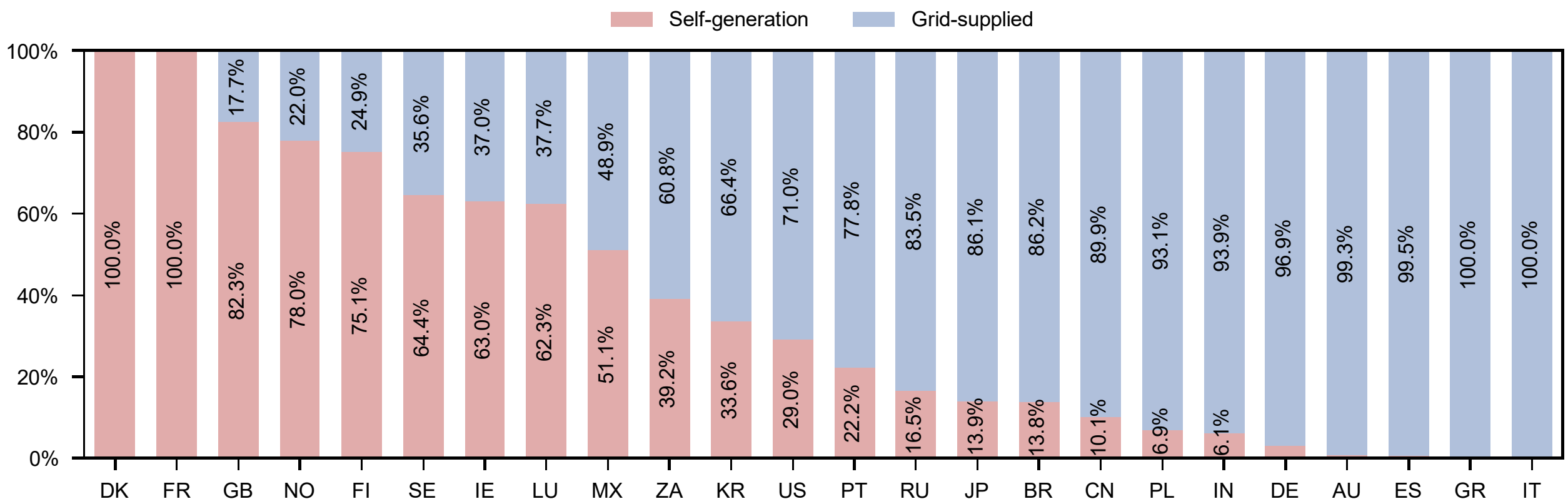

**Appendix Fig. 4 Deployment model of data centres across major countries.**

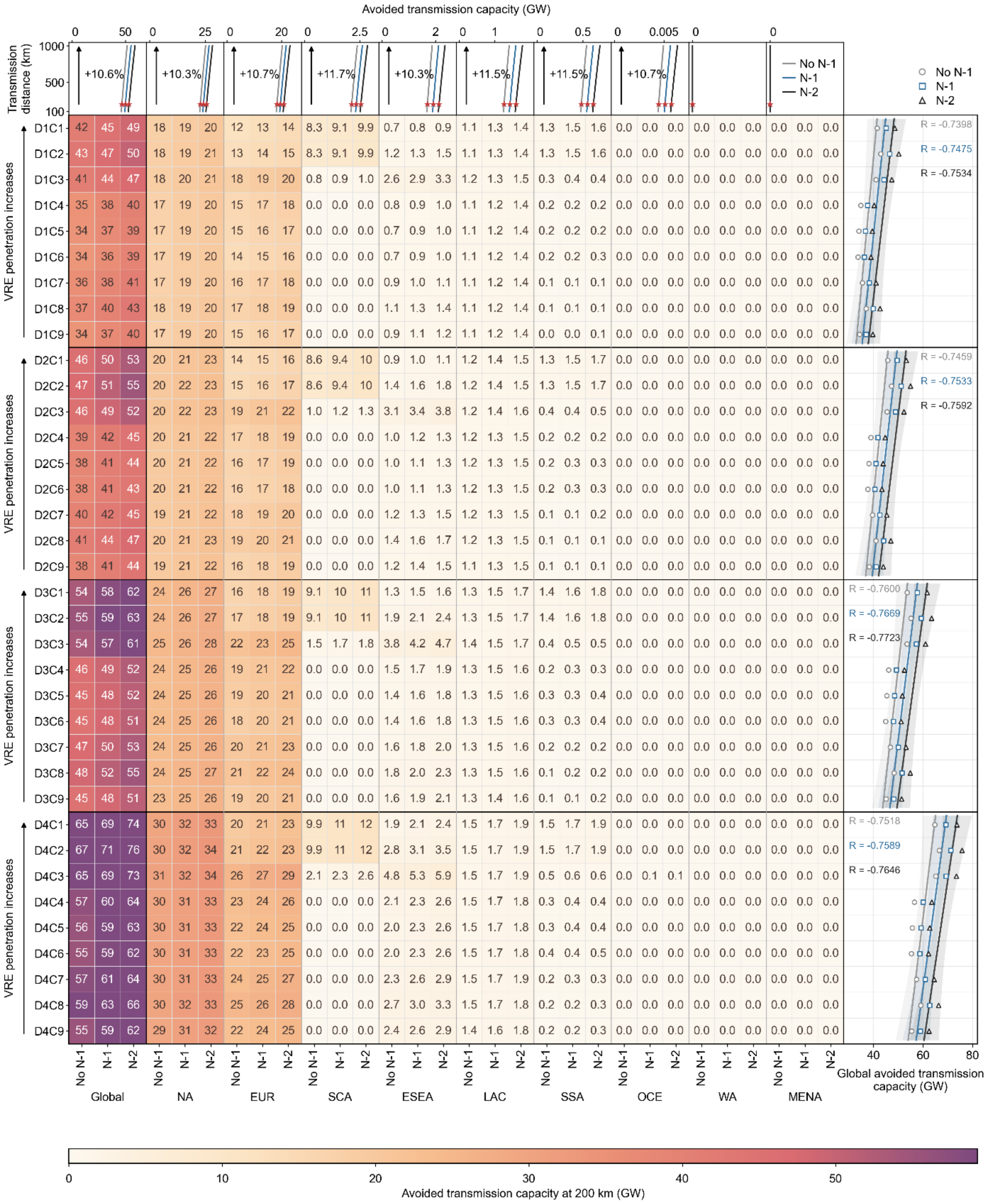

**Appendix Fig. 5 Avoided transmission capacity under computation-transfer-based data-centre siting.** The heatmap reports avoided transmission capacity at a 200-km equivalent transmission distance across different scenarios, under three grid-reliability assumptions: no N-1 constraint, N-1 and N-2. Upper plots show sensitivity to equivalent transmission distances from 100 to 1,000 km. The right-hand panel relates global avoided transmission capacity to VRE penetration, with points denoting individual scenarios and lines showing fitted trends.